\documentclass[twocolumn,twocolappendix]{aastex63}
\pdfoutput=1 
\usepackage{amsmath,amstext}
\usepackage[T1]{fontenc}
\usepackage[figure,figure*]{hypcap}
\usepackage{booktabs}
\usepackage{graphicx}

\newcommand{\tess}{{\it TESS}}
\newcommand{\kep}{{\it Kepler}}
\newcommand{\ktwo}{{\it K2}}

\newcommand\pchi{\ifmmode{P(\chi^2)}\else $P(\chi^2)$\fi}
\newcommand\kms{\ifmmode{\rm km\thinspace s^{-1}}\else km\thinspace s$^{-1}$\fi}
\newcommand\ms{\ifmmode{\rm m\thinspace s^{-1}}\else m\thinspace s$^{-1}$\fi}
\newcommand\msun{\ifmmode{M_{\odot}}\else $M_{\odot}$\fi}
\newcommand\rsun{\ifmmode{R_{\odot}}\else $R_{\odot}$\fi}
\newcommand\lsun{\ifmmode{L_{\odot}}\else $L_{\odot}$\fi}
\newcommand\mstar{\ifmmode{M_{\star}}\else $M_{\star}$\fi}
\newcommand\rstar{\ifmmode{R_{\star}}\else $R_{\star}$\fi}
\newcommand\mjup{\ifmmode{M_{\rm Jup}}\else $M_{\rm Jup}$\fi}
\newcommand\rjup{\ifmmode{R_{\rm Jup}}\else $R_{\rm Jup}$\fi}
\newcommand\mearth{\ifmmode{M_\oplus}\else $M_\oplus$\fi}
\newcommand\rearth{\ifmmode{R_\oplus}\else $R_\oplus$\fi}
\newcommand\mpl{\ifmmode{M_{\rm P}}\else $M_{\rm P}$\fi}
\newcommand\rp{\ifmmode{R_{\rm P}}\else $R_{\rm P}$\fi}
\newcommand\vsini{\ifmmode{v\sin{i_\star}}\else $v\sin{i_\star}$\fi}
\newcommand\sini{\ifmmode{\sin{i_\star}}\else $\sin{i_\star}$\fi}
\newcommand\prot{\ifmmode{P_{\rm rot}}\else $P_{\rm rot}$\fi}
\newcommand\logg{\ifmmode{\log{g}}\else $\log{g}$\fi}
\newcommand\teff{\ifmmode{T_{\rm eff}}\else $T_{\rm eff}$\fi}

\newcommand\mysim{\mathord{\sim}}

\def\lsim{\mathrel{\rlap{\lower4pt\hbox{\hskip1pt$\sim$}}
    \raise1pt\hbox{$<$}}}
\def\gsim{\mathrel{\rlap{\lower4pt\hbox{\hskip1pt$\sim$}}
    \raise1pt\hbox{$>$}}}

\newcommand{\rffigl}[1]{Figure~\ref{fig:#1}}

\newcommand{\rfsecl}[1]{\mbox{$\S$\ref{sec:#1}}}

\newcommand{\rftabl}[1]{Table~\ref{tab:#1}}

\shortauthors{Paredes et al.}

\begin{document}

\title{The Solar Neighborhood XLIX: Nine Giant Planets Orbiting Nearby K Dwarfs, and the CHIRON Spectrograph's Radial Velocity Performance}

\author[0000-0003-1324-0495]{Leonardo A. Paredes}
\affiliation{Department of Physics and Astronomy, Georgia State University, Atlanta, GA 30302-4106, USA} 

\author[0000-0002-9061-2865]{Todd J. Henry}
\affiliation{RECONS Institute, Chambersburg, PA 17201, USA}

\author[0000-0002-8964-8377]{Samuel N. Quinn}
\affiliation{Center for Astrophysics, Harvard and Smithsonian, 60 Garden Street, Cambridge, MA 02138, USA}

\author[0000-0001-8537-3583]{Douglas R. Gies}
\affiliation{Department of Physics and Astronomy, Georgia State University, Atlanta, GA 30302-4106, USA}

\author{Rodrigo Hinojosa-Goñi}
\affiliation{Cerro Tololo Inter-American Observatory, CTIO/AURA Inc., La Serena, Chile}

\author[0000-0003-4568-2079]{Hodari-Sadiki James}
\affiliation{Department of Physics and Astronomy, Georgia State University, Atlanta, GA 30302-4106, USA}

\author[0000-0003-0193-2187]{Wei-Chun Jao}
\affiliation{Department of Physics and Astronomy, Georgia State University, Atlanta, GA 30302-4106, USA}

\author[0000-0001-5313-7498]{Russel J. White}
\affiliation{Department of Physics and Astronomy, Georgia State University, Atlanta, GA 30302-4106, USA}

\accepted{\it 2021 June 2}
\published{\it 2021 October 4}
\submitjournal{\it The Astronomical Journal}


\begin{abstract}

We report initial results of a large radial velocity survey of K dwarfs up to a distance of 50 pc from the Solar System, to look for stellar, brown dwarf, and jovian planets using radial velocities from the CHIRON spectrograph on the CTIO/SMARTS 1.5m telescope. We identify three new exoplanet candidates orbiting host stars in the K dwarf survey, and confirm a hot Jupiter from \tess\ orbiting TOI 129. Our techniques are confirmed via five additional known exoplanet orbiting K dwarfs, bringing the number of orbital solutions presented here to 9, each hosting an exoplanet candidate with a minimum mass of 0.5--3.0 $\mjup$. In addition, we provide a list of 186 nearby K dwarfs with no detected close companions that are ideal for more sensitive searches for lower mass planets. This set of stars is used to determine CHIRON's efficiency, stability, and performance for radial velocity work. For K dwarfs with V = 7--12, we reach radial velocity precisions of 5--20 \ms~under a wide range of observing conditions. We demonstrate the stability of CHIRON over hours, weeks, and years using radial velocity standards, and describe instrumental capabilities and operation modes available for potential users.

\end{abstract}


\section{Introduction}

The field of exoplanet discovery has advanced quickly, from the first detections of planets \citep{latham:1989,wolszczan:1992,mayor:1995}, to large-scale transit surveys from space transits such as {\it Kepler} \citep{borucki:2010}, {\it CoRoT} \citep{baglin:2006}, and \tess\ \citep{ricker:2014}. This paper is concerned primarily with large scale radial velocity surveys targeting 1000 stars or more, such as the Lick/Keck survey \citep{butler:2017} and the CORAVEL/HARPS survey \citep{mayor:2011}, that provide key statistics about the formation of stellar and planetary systems. Such large surveys enable ensemble studies to complement individual discoveries, but are also valuable in the evaluation of observing facilities, techniques, and results. In this spirit, here we provide results from the CHIRON spectrograph on the CTIO/SMARTS 1.5m \citep{tokovinin:2013}, which is being used as part of a multi-faceted survey of more than 5000 of the nearest K dwarfs.

High resolution spectrographs used for precise radial velocity work are critical components of exoplanet surveys, and are particularly relevant now due to their symbiosis with transit survey instruments on both the ground and in space that are used to detect and characterize planetary systems. In the southern hemisphere there have been at least eight spectrographs precise enough to detect giant planets, operated in various mixtures of classical and queue observations, with or without assistance from the instruments' Principal Investigators (PIs).
These instruments include HARPS \citep{mayor:2003} at ESO La Silla 3.6m Telescope, FEROS \citep{kaufer:1998}, CORALIE (former ELODIE) \citep{baranne:1996} at ESO La Silla Swiss 1.2-metre Leonhard Euler Telescope, PFS \citep{crane:2006} and MIKE \citep{bernstein:2003} at Las Campanas Observatory 6.5m Magellan II Telescope, FIDEOS \citep{tala:2014} at ESO La Silla 1-metre Telescope, and Veloce Rosso \citep{gilbert:2018} at the Anglo-Australian Telescope. In this paper, we describe initial science results from our radial velocity search for companions orbiting the nearest K dwarfs carried out at the Cerro Tololo Inter-American Observatory (CTIO) in Chile, and provide details about CHIRON operations that are useful for those searching for exoplanets, driven in particular by NASA's \tess\ mission.

Since 1994, the RECONS group has been studying the nearest stars \citep{henry:1997}, with various scientific investigations focused on horizons spanning 10 to 100 pc. One of the key results is that the population of stars in the solar neighborhood is dominated by stars smaller than the Sun, with M dwarfs accounting for 75\% of all stars, followed by K dwarfs at 12\% (\citet{henry:2006, henry:2018} with updates at {\it www.recons.org}). The K dwarfs are the focus of this paper, as they lie in a sweet spot in between the shorter-lived, rarer G dwarfs (5\% of the population) and the magnetically active M dwarfs, making them arguably the most suitable hosts for long-term, biologically active planets \citep{cuntz:2016}.

In this paper we discuss first results of a survey of several hundred K dwarfs using the CHIRON instrument on the CTIO/SMARTS 1.5m telescope. CHIRON is being used as part of NASA's efforts to follow-up \tess\ exoplanet detections and for individual PI spectroscopic surveys of nearby K and M dwarfs. We provide key statistics on the capabilities of CHIRON for K dwarfs with magnitudes of $\sim$8--11 using a set of presumably single stars, resulting in typical radial velocity precisions of 5--20 \ms. In addition, we present results for giant planets detected orbiting K dwarfs, including (a) five previously known planets used to check our observing protocols, reduction techniques, and overall CHIRON performance, (b) a detection of a planet orbiting NLTT 58744 (here after HIP 65) from \tess\ (TOI 129), and (c) three candidate planets orbiting the K dwarfs HIP 5763, HIP 34222, and HIP 86221 in our larger survey.

\section{CHIRON Operations}

The CTIO/SMARTS 1.5m telescope has been operated by the SMARTS Consortium \citep{subasavage:2010} since 2003. The telescope is primarily operated by onsite CTIO/SMARTS staff observers, although occasional runs are done by individual SMARTS Consortium members. The 1.5m has had a few different instruments available since SMARTS began, but the primary instrument over the past decade, and the only one operating since 2015 is the CHIRON high-resolution spectrograph \citep{tokovinin:2013}. CHIRON is operational since in 2011, it was upgraded in 2012 and has been used until the 1.5m closed in 2016, when the primary observing program at the time ended. The 1.5m was reopened in June 2017 via an effort by RECONS team members Paredes and Henry working closely with CTIO staff members, and the 1.5m has now been operating exclusively with CHIRON for three years.

\subsection{Observing Queue Management}

Since June 2017, the RECONS Team has overseen the management and operations of the 1.5m and CHIRON. This includes scientific support, the construction of observing queues, data processing, and data distribution. Queue management is carried out via a web-based platform \href{chiron.astro.gsu.edu}{chiron.astro.gsu.edu} in which time allocations, target requests, desired observing cadences, and on-site telescope operations are fully integrated. Programs from users who purchase observing time and those awarded time via the NOAO/NOIRLab and Chilean TACs are scheduled simultaneously.

Once observing time is allocated, PIs and their collaborators access the queue management platform to submit their targets and observing requirements, managing the use of the observing time that they have been granted. Our queue management team then sorts the various observing requests from multiple PIs to create the nightly schedule, and the observing sequence is executed by a CTIO/SMARTS observer using the same web-based platform. The observer is able to add, remove, or re-sequence observations as sky conditions permit, and can complete the arc of some science programs through a 7-night shift or prepare the next shift's observer to carry on longer programs. From October 2017 through July 2019, the facility was operated every other week, while beginning in August 2019, operations expanded to full-time nightly coverage.

\subsection{Telescope Operations}

A typical night of CHIRON operations starts with a fixed set of calibrations for all observing modes, detailed in \rftabl{slitmodes}, that commence in the afternoon. At astronomical twilight, the on-site observer accesses the night's observing schedule using the web platform and executes the plan, reporting back the status and/or any issues that may affect the observations. For some programs, an Atlanta-based team member at Georgia State University is available real-time to assist in interpretation of the science requirements. The web platform is directly connected with the instrument controller; therefore, on each target acquired the requested instrument setup is passed directly and efficiently to the instrument and telescope control system (TCS), minimizing time spent and configuration errors. At the end of the night, another fixed set of calibrations is secured. More details are explained in \cite{brewer:2014}. 

\subsection{Data Reduction and Distribution}

The raw data and calibration files are backed up and transferred for processing daily to computer facilities in Atlanta. The files consist of 2D CCD frames containing the echelle orders and header information, where telescope, ephemerides, spectrograph, and exposure meter (EM) data are recorded. The default data processing consists of bias and flat-field corrections, cosmic ray removal, echelle order extractions, and wavelength calibrations using ThAr comparison lamps.\footnote{An iodine cell is available, but few CHIRON PIs have used \\ the iodine cell since reopening and wavelength calibrations \\ based on iodine lines are not part of routine operations.}  The algorithms used are based on the REDUCE IDL package \citep{piskunov:2002} and were adapted to CHIRON data with the application of deriving precise radial velocity measurements. Reduced data are produced in FITS file format, containing for each extracted order the flux in photo-electrons and topocentric wavelength in Angstroms per pixel. Additional specifics of the data reduction process are described in detail in \cite{tokovinin:2013}. Once a night of observations is fully processed, calibration files, raw data, logs, and reduced data are grouped into individual PI programs and placed in secure directories on servers in Atlanta. PIs are then notified and instructed how to retrieve their data products. Under normal operating conditions, the entire process from raw data acquisition to the delivery of fully processed data products is completed within few days.

\section{CHIRON Capabilities}

Over the years, the CHIRON spectrograph on the 1.5m has been used to pursue a wide variety of scientific goals, but it was envisioned as an instrument to measure radial velocities precise enough to detect planets orbiting bright stars \citep{giguere:2015}. Other science goals accomplished with CHIRON data have involved optical spectral characterization of a variety astrophysical objects, such as the atmospheres of bright stars, novae \citep{munari:2016,giguere:2016}. The results we present here aim to describe the performance capabilities of CHIRON for measuring radial velocities.

\subsection{Specifications and Setup Options}
\label{sec:specsandsetup}

CHIRON is able to cover an optical wavelength range between 415 and 880 nm divided in spectral orders from 136 to 66, and is able to acquire targets up to $\sim$18 mag through a fiber that is 2.7\arcsec in diameter in the sky. The primary resource for details about CHIRON is \cite{tokovinin:2013}, which outlines the four different slit setups available, ranging in resolution from 28 000 to 136 000 (details given in \rftabl{slitmodes}). A user's choice of setup is dependent on science goals and target brightnesses.

\begin{deluxetable}{lcccc}
\setlength{\tabcolsep}{0.032in}
\tabletypesize{\small}
\tablecaption{\label{tab:slitmodes} Modes available for observing with CHIRON.}
\tablehead{\colhead{Slit Mode}    & 
           \colhead{Binning}      & 
           \colhead{R}            & 
           \colhead{Throughput}   &
           \colhead{Element Size}}
\startdata
Fiber       & 4x4 &  28 000  & 1.00  & 4000 \ms   \\
Slicer      & 3x1 &  80 000  & 0.75  & 1000 \ms   \\
Slit        & 3x1 &  90 000  & 0.40  & 1000 \ms   \\
Narrow Slit & 3x1 & 136 000  & 0.20  & 1000 \ms   \\
\enddata
\tablecomments{The K dwarf survey described in this paper uses slicer mode.}
\end{deluxetable}

CHIRON offers two wavelength calibration options --- a ThAr comparison lamp and an iodine cell. The latter can be used to achieve instrumental precision below 5 \ms~on targets brighter than $V \sim 6$, at the expense of requiring large amounts of observing time to reach a sufficient signal-to-noise ratio (S/N) for such precision. The work we present here utilizes the ThAr lamp for wavelength calibration, which is more versatile than the iodine cell and is the most popular among users.

\subsection{CHIRON Setup for the K Dwarf Program}
\label{sec:setup4survey}
The K dwarfs are being observed with CHIRON in slicer mode, which provides a resolution of 80000 and spreads the spectrum into 59 orders. Integration times are uniformly set to 900 seconds except for a few stars brighter than $V\sim$ 6 for which the exposure is typically stopped when the S/N reaches 100 at 5500\AA. After each science exposure, a single ThAr lamp exposure of 0.4 seconds is taken to use for wavelength calibration. In poor seeing or partial cloud cover, the 900 second integration time is maintained in an effort to permit coverage of targets in the large program and to provide insight into how precision changes under various sky conditions. The consistent 900 second integration times for stars with $V \sim$ 6--12 also enables straightforward evaluations of the changes in RV precision with magnitude, as well as permitting direct comparisons of fluxes received for individual stars throughout their coverage during the survey.

\subsection{Radial Velocity Pipeline}
\label{sec:rvpipeline}

A pipeline (here after the RV pipeline) was developed in python by this group to process uniformly large volumes of spectra taken with CHIRON to extract radial velocities. It was optimized specifically for our K dwarf survey at the 1.5m and follows a recipe that has nine steps.

\begin{enumerate}
\item {\it Input spectra.} Each wavelength-calibrated spectrum (\rffigl{inputspectra}) enters the RV pipeline once it has been confirmed to have the proper target identification, as well as correct observational parameters, such as coordinates, hour angle, time of exposure, and airmass.

\begin{figure}[ht!]
    \centering
    \includegraphics[width=0.47\textwidth]{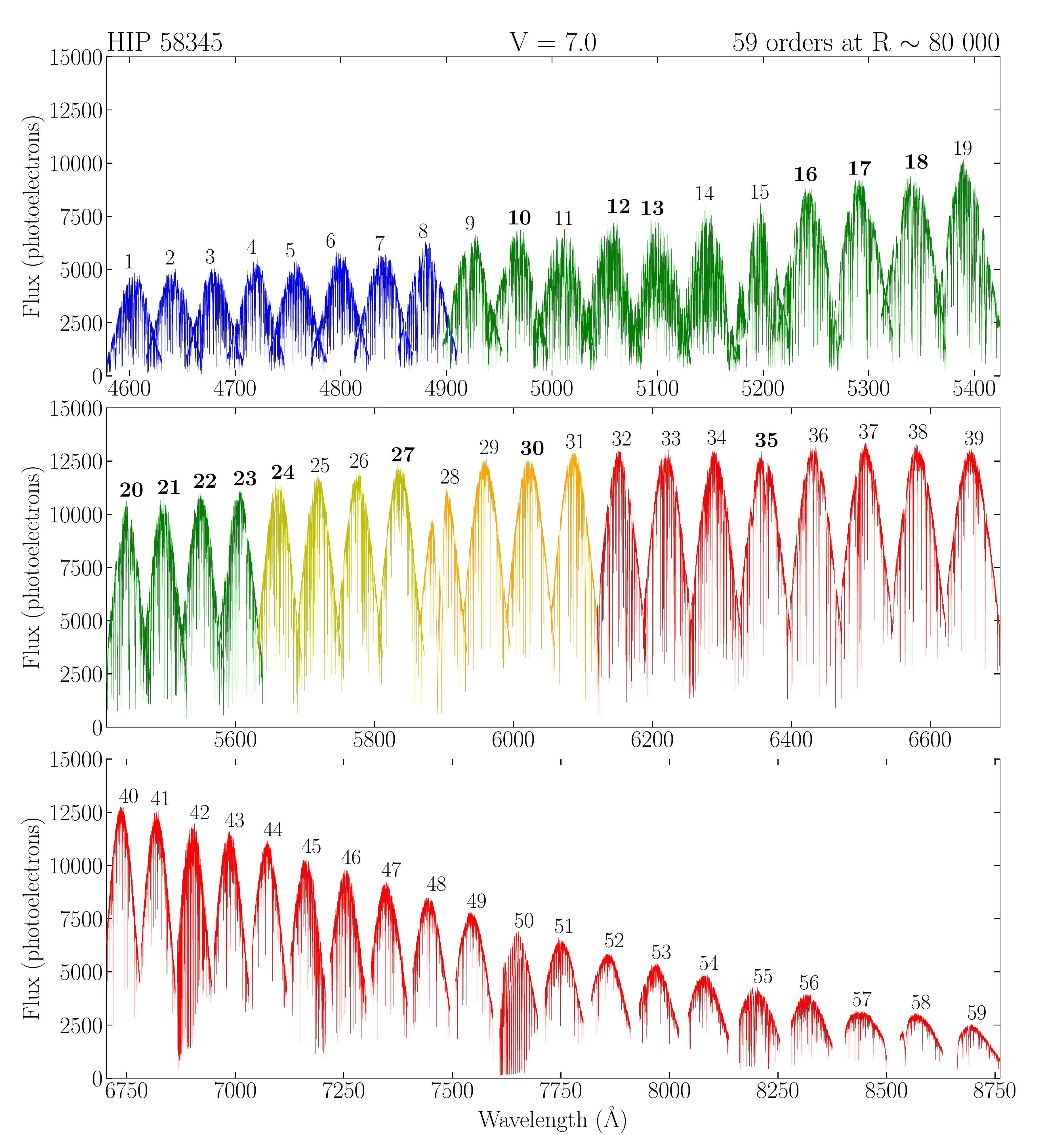}
    \label{fig:inputspectra}
    \caption{59 echelle spectral orders in slicer mode (R $\sim$ 80000) for the K dwarf HIP 58345 extracted by the CHIRON basic reduction pipeline. A fixed set of 14 of these orders (highlighted) are used to derive our radial velocities.}
\end{figure}

\item {\it Flattening spectra.} The removal of the blaze function embedded on each order of echelle spectra is crucial for extracting precise radial velocities. The Doppler information contained in the spectrum is most valuable in regions where changes in the slope of the flux per wavelength are steepest, i.e., for sharp lines. The flattening process starts with a recursive sigma clipping algorithm to remove the spectral lines over sub sections of the order to be able to map the continuum. Subsequently, a 5th-order polynomy is fitted using remaining data points of the order. Finally, the original unfiltered spectral order is divided by the the fitted blaze function to get the flattened normalized version (\rffigl{blazeremoval}). Removing the blaze function to flatten each order ensure that Doppler shifts are the result of shifting lines due to a star's velocity, rather than from changes in the instrument response along the order. \rffigl{flatspectra} illustrates three spectral regions after removing the blaze function and normalization. These spectral regions include many key features found in K dwarf spectra: \ion{Cr}{1}, \ion{Fe}{1}, H$\alpha$, \ion{Li}{1}, \ion{Na}{1}, and \ion{Ti}{1}.

\begin{figure}[ht!]
    \centerline{\includegraphics[width=0.47\textwidth]{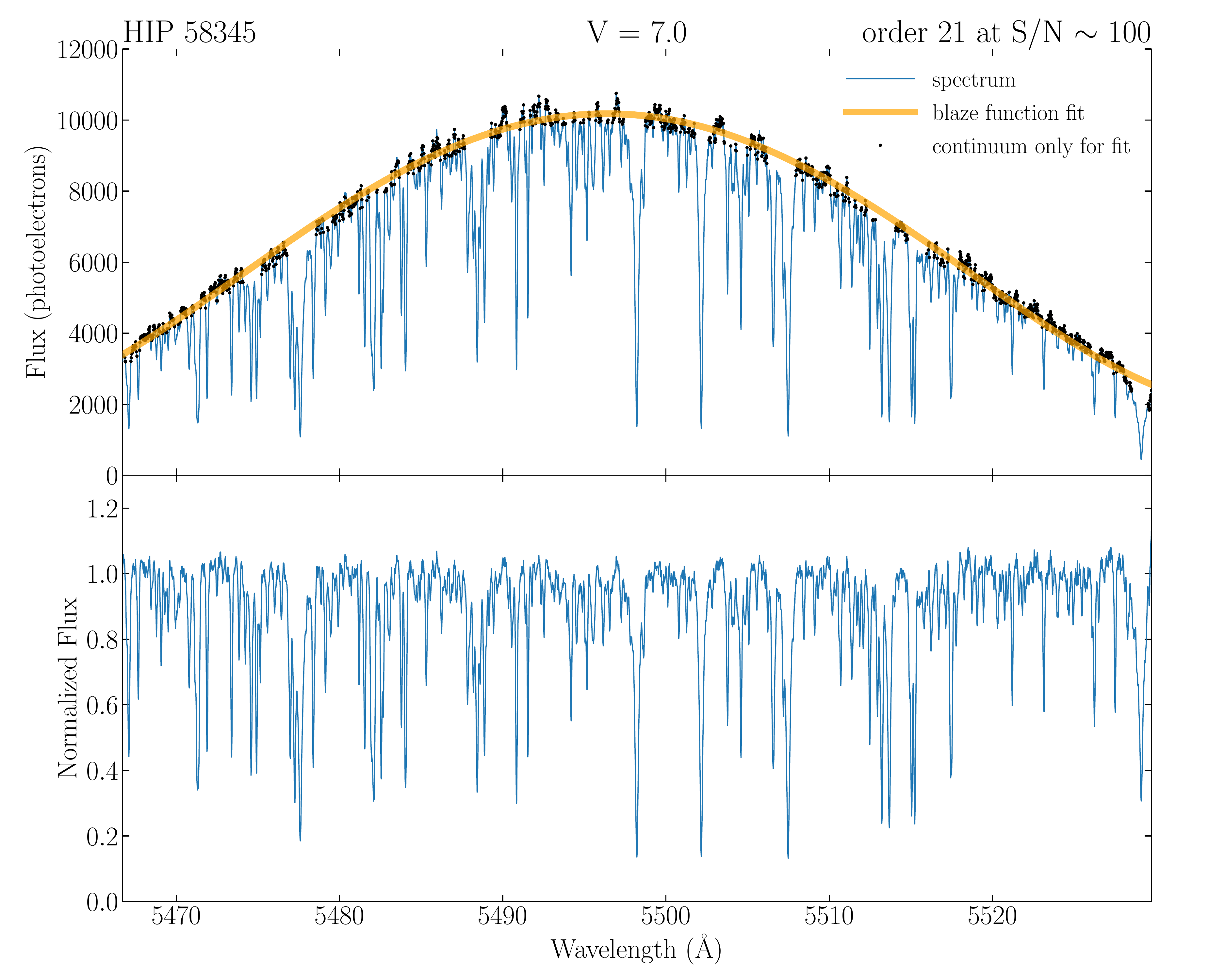}}
    \label{fig:blazeremoval}
    \caption{Spectrum of order 21 ($\sim$5460 to 5530 \AA) of HIP 58345 taken in slicer mode (R $\sim$ 80 000) at S/N $\sim$ 100. \textit{top:} spectrum before the removal of the blaze function, where the black dots are the data points from the original spectrum (blue line) used to fit the blaze function (yellow line). \textit{bottom:} Normalized spectrum after the removal of the blaze function.}
\end{figure}

\begin{figure}[ht!]
    \centerline{\includegraphics[width=0.47\textwidth]{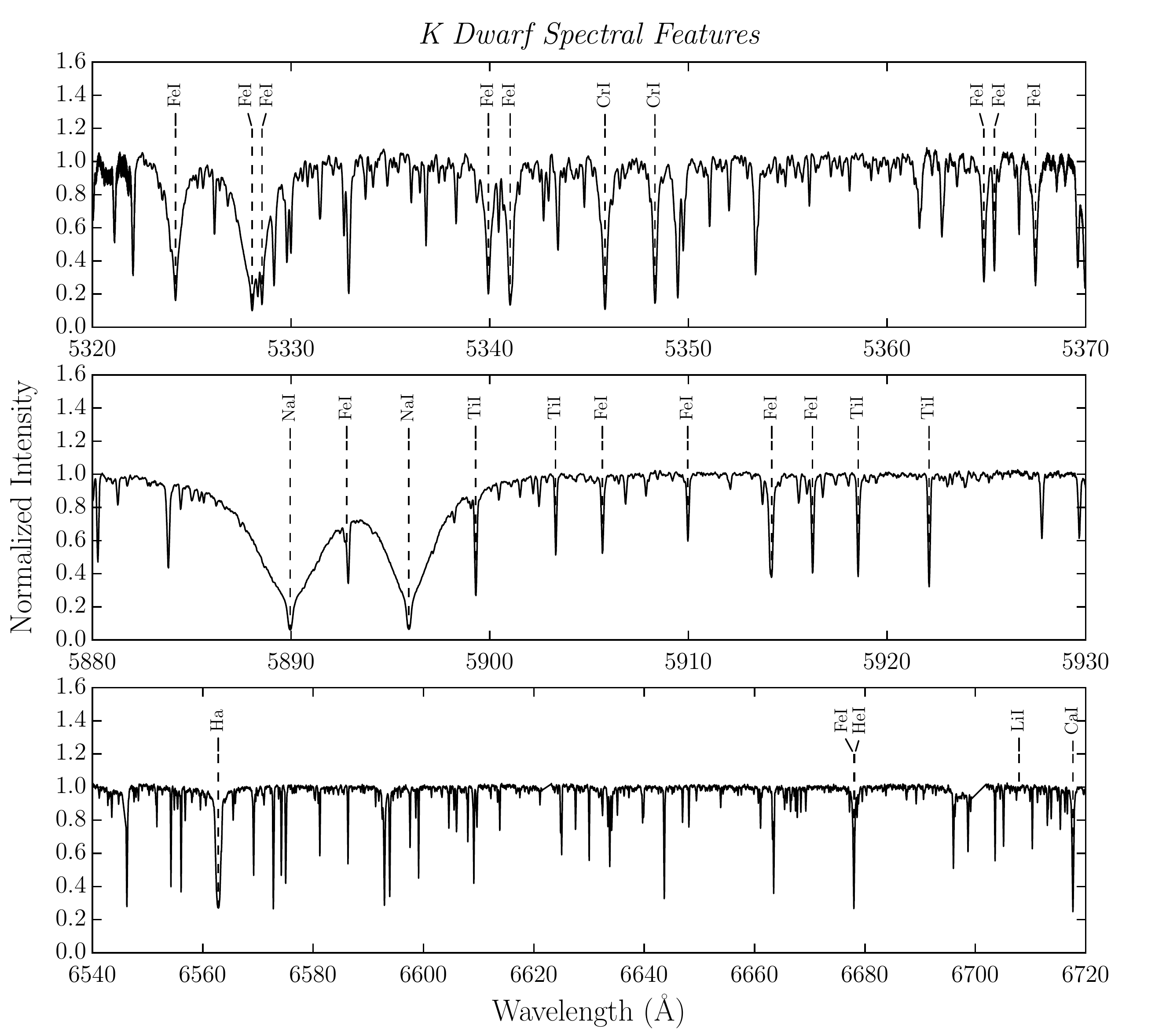}}
    \label{fig:flatspectra}
    \caption{Examples of three spectral regions of the K dwarf HIP 73184 observed with CHIRON, after removing the blaze function and normalizing each order.}
\end{figure}

\item {\it Barycentric correction.} The motion of the Earth around the barycenter of the Solar System produces a large ($\pm$30 \kms) Doppler oscillation present in all sequences of spectra. These variations must be removed with high precision to derive the final radial velocities for a targeted star. We use the algorithm "barycorr" by \cite{wright:2014}, which calculates corrections appropriate for radial velocity precisions to better than 3 \ms. The three ingredients used to calculate the correction are (1) the geographical position of the CHIRON spectrograph on Earth, using GPS measurements by \cite{mamajek:2012}, (2) the time stamp of the observation, taken as the midpoint of the exposure weighted by photon counts as measured via the exposure meter of CHIRON, saved in the image header under keyword EMMNWOB, and (3) astrometric information for the target star, including its RA, DEC, proper motion and parallax. For each spectrum a barycentric velocity correction in \ms\ is obtained to be subtracted from the radial velocity derived , and a barycentric julian date in days is obtained that becomes the time-stamp for the corrected radial velocity.

\item {\it Choose template spectrum.} We calculate a radial velocity at each epoch relative to a single epoch observation we call the template spectrum for each star. We select the best quality spectrum as template, considering weather conditions, Moon illumination and S/N.

\item {\it Re-sample spectrum.} The spectrum is re-sampled to match the same wavelength grid as the template spectrum to enable direct positional comparisons for the cross-correlation matches. Once the template spectrum is selected, each order is interpolated into a linear log-wavelength grid and oversampled two times the total number of pixels (6400 pixels per order). The result is that any single pixel Doppler shift across the spectrum corresponds to a radial velocity shift of 500 \ms~in the slicer mode used for this survey.

\item {\it Order selection.} One of the advantages of working with spectra from a single spectral type and luminosity class is that spectral features do not vary significantly from one star to another. Therefore, we select a set of spectral orders out of the full set of 59 orders provided by CHIRON on slicer mode that provides the best Doppler results. The quality of the Doppler information given within a single order depends on the number of spectral lines present, the shapes of those lines, the prevalence of telluric lines that pollute the spectrum, and the S/N across the order. Better precision in the radial velocity calculation comes from orders with more numerous spectral lines that are sharp and deep, with few telluric lines, and with less noise. Following these criteria, we omit orders at the extreme ends of the total wavelength range that have relatively low S/N --- these are also furthest in wavelength scale from the instrument's peak efficiency at 5500 \AA, as well as being relatively far from the blackbody peaks of K dwarfs. In addition, some orders were omitted because they simply have too few lines or are severely contaminated by telluric lines. Finally, orders were removed that have sources of contamination to the radial velocity signal resulting from broad spectral lines and lines sensitive to stellar activity, such as H$\alpha$, H$\beta$, and the Na doublet. In the end, we use a set of 14 orders selected from the 59 available in slicer mode: 10, 12, 13, 16, 17, 18, 20, 21, 22, 23, 24, 27, 30, 35. Note that these are arbitrary order number labels assigned for the reduced data products of CHIRON, where the central wavelength of order $n$ in \AA\ is given by $\lambda_n = 565754/(124-n)$.

\item {\it Cross-correlation function.} The radial velocity at a given epoch is calculated relative to the epoch of the selected template spectrum. The wavelength grid of each spectrum is matched to the template spectrum, and then the cross-correlation function (CCF) is calculated for each order pair. The radial velocity derived from the order corresponds to the location of peak of the CCF, for which the location is determined by fitting a Cauchy-Lorentz function. Additional parameters of the fitted function such as FWHM and amplitude are also obtained and used for the uncertainty on the  estimated RV. 

\item {\it Uncertainty estimation.} The uncertainty of the radial velocity extracted from each order is closely related to the criteria described in step (6), in concert with the resulting shape of the derived CCF. We estimate a velocity uncertainty for a single order following the prescription by \cite{zucker:2003} by quantifying the relative amplitude and sharpness of its CCF. The CCF shape and quality are directly related to the S/N of the spectrum; therefore, the errors estimated this way are photon errors. Instrumental errors and astrophysical noise are not reflected in this value, so the quoted errors may underestimate the total uncertainty.

\item {\it Radial velocity calculation.} Once each order from a given spectrum is cross-correlated with its respective order in the template, a final radial velocity and its uncertainty are computed as in step 7. The final value and uncertainty for the epoch's observation is derived using the individual values and their uncertainties from the 14 orders by determining a weighted mean value and the standard error on the weighted mean. 
\end{enumerate}

\subsection{CHIRON Stability}

\begin{figure}
\vspace{0.4in}
\centerline{\includegraphics[width=\linewidth]{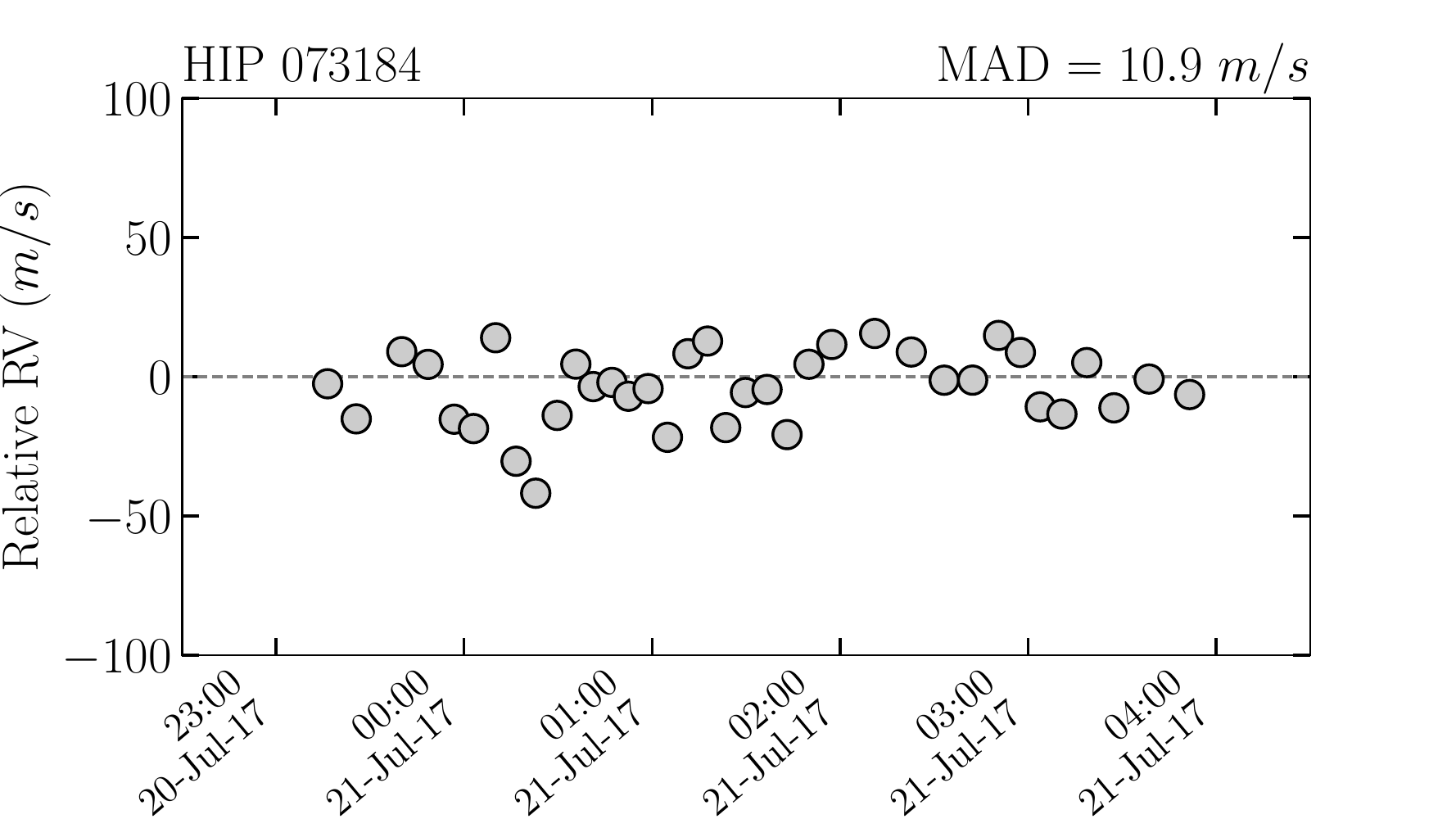}}
\centerline{\includegraphics[width=\linewidth]{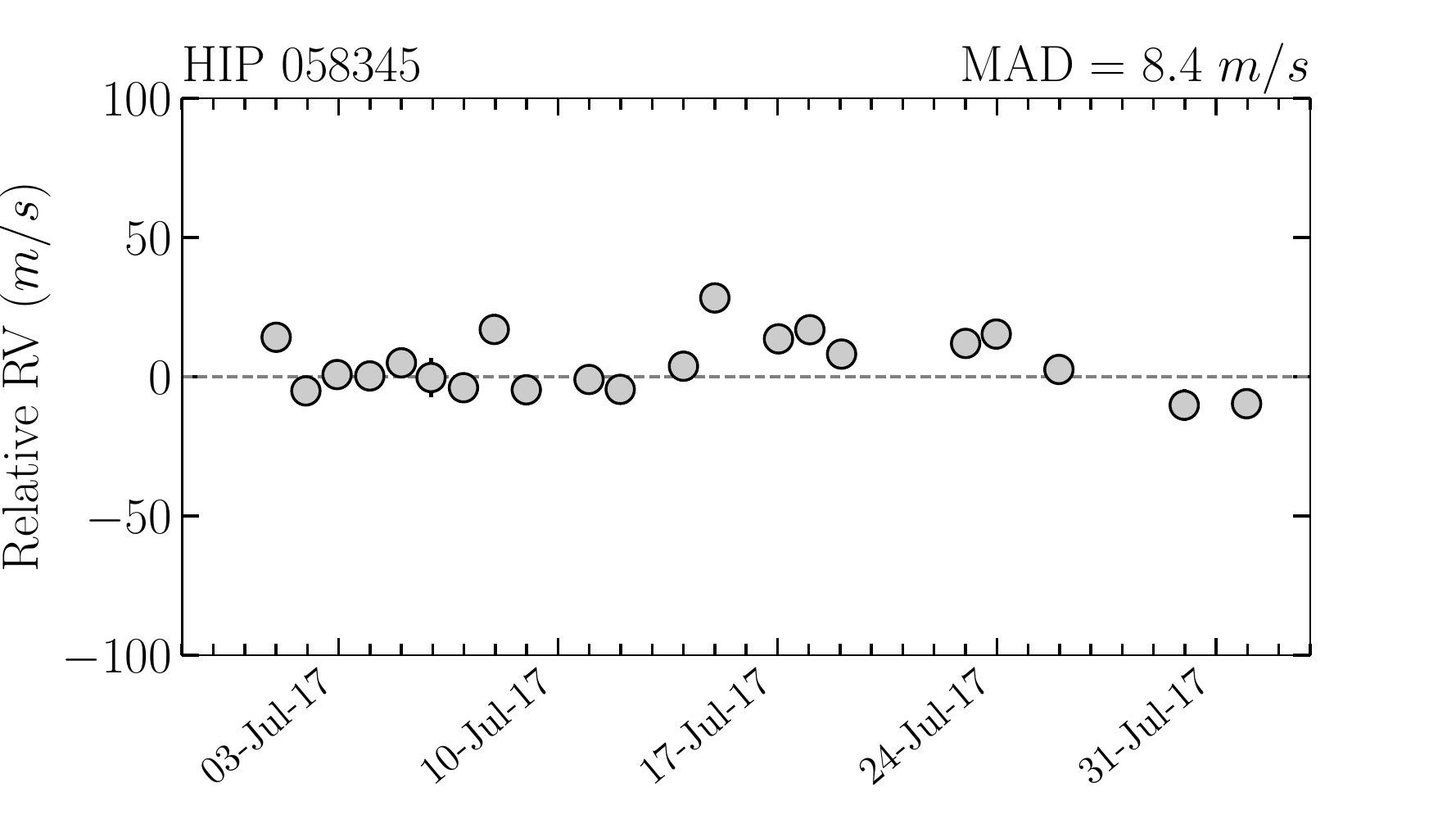}}
\centerline{\includegraphics[width=\linewidth]{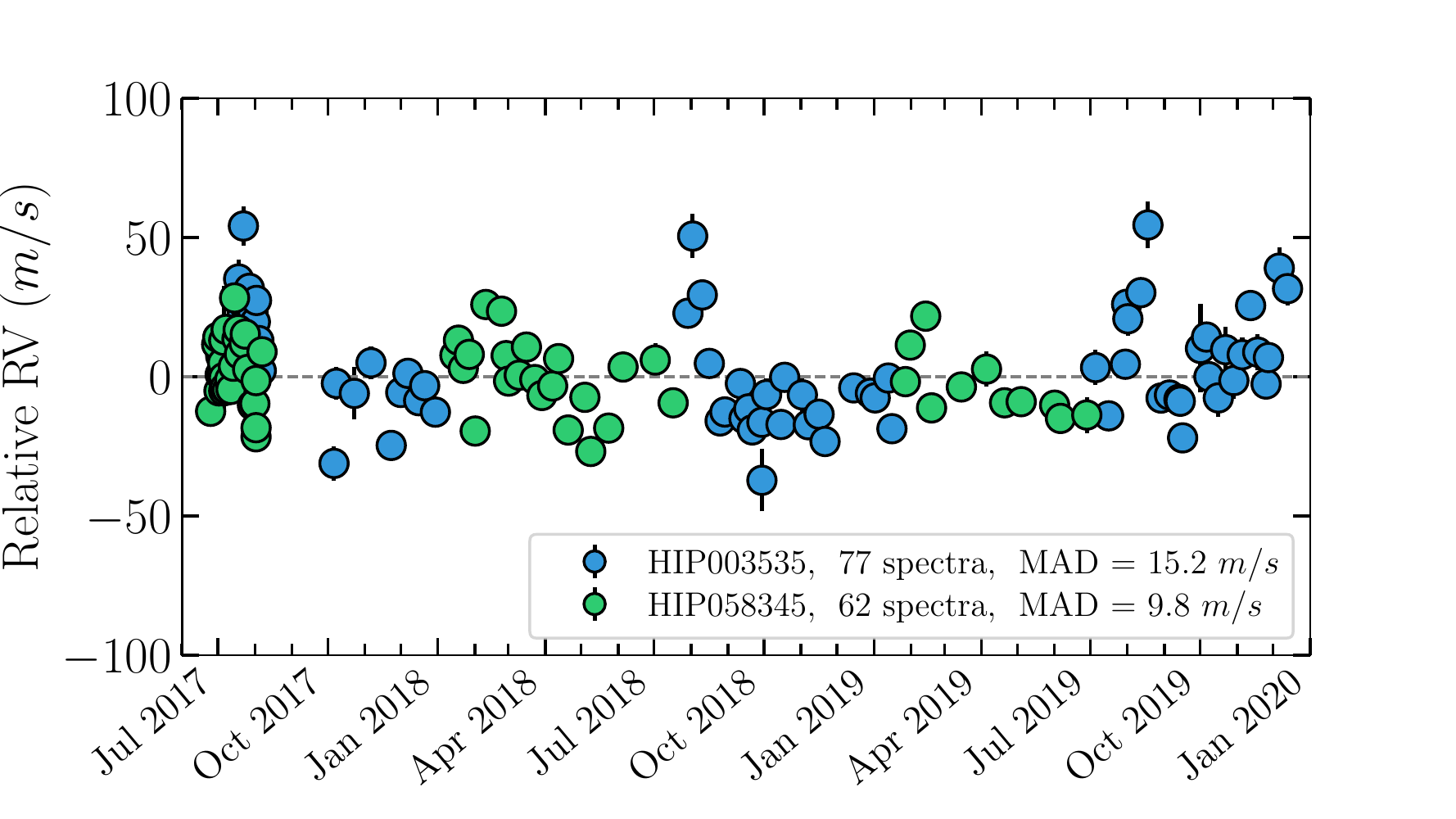}}
\vspace{0.3in}
\label{fig:standards}
\caption{Three K dwarfs used as RV standards to monitor the stability of CHIRON. \textit{Top}: HIP 73184 observed for $\sim$4.5 hr on a single night. \textit{Middle}: HIP 58345 observed roughly once per night for a month; skipped nights were due to unfavorable weather conditions. \textit{Bottom}: HIP 3535 and HIP 58345 observed between 2017 June and 2019 December with a typical cadence of one observation every 7--10 days.}
\end{figure}

Here we provide details about the stability of CHIRON on three timescales: over a night, a month, and the more than two years since the 1.5m was reopened. Three K dwarf standard stars, HIP 3535 ($V$ = 8.0), HIP 58345 ($V$ = 7.0), and HIP 73184 ($V$ = 5.8), have been observed since the telescope was reopened in June 2017, and combined, two (HIP 3535 and HIP 58345) now provide a consistent stream of data as they compensate for one another's seasonal gaps. These three stars were selected because they had data from previous RV programs indicating that they have variations of only 3--7 \ms~over years \citep{butler:2017}, sufficiently low for our purposes to consider them RV standard stars. Results for all three stars are shown in Figure \ref{fig:standards} and with RV data listed in Table \ref{tab:rvstds}. There are occasional outlier points, in particular for HIP 3535, the faintest of the three standards.

The error associated to each radial velocity measured on a night is tightly correlated with the photon flux received by the CHIRON detector and relates to the variance across the 14 orders for which individual RVs are extracted, but as can be seen in the time series for HIP 3535, for example, the dispersion as measured by the mean absolute deviation (MAD) of the points overall (15 \ms) is larger than errors on most of the individual points (typically 3--10 \ms). The roughly factor of two difference is presumably due to systematic errors, with the leading culprits being changes in the focus of lines on the CCD and temperature fluctuations inside the chamber that houses the CHIRON instrument.

Once multiple spectra are available for a given star, the MAD is calculated for all available spectra, which is $\sim$0.8 times the standard deviation of normally distribute data. It is the MAD values for various stars that are given in the panels of \rffigl{standards} and reported henceforth in this paper. We emphasize that these results are for K dwarfs observed during our survey; other types of stars will not necessarily provide similar results, i.e., hotter stars with fewer lines available for RV extraction, or rapidly rotating stars with broad lines.

\begin{enumerate}
\item {\it Stability over hours.} HIP 73184 was observed 36 times on the night of 2017 July 17 to test the stability of CHIRON over a period of $\sim$4.5 hours. Observations were taken on a clear night with seeing of 0.7--1.0\arcsec, while the airmass changed from 1.012--1.674. The top panel of \rffigl{standards} shows that the MAD is 10.9 \ms\ during the series of observations.

\item {\it Stability over 1 month.} As shown in middle panel of \rffigl{standards}, HIP 58345 was observed on 21 nights over a one month period in 2017. There are few RV measurements lying far from the mean value, with a resulting MAD of 9.8 \ms~for the data series. This MAD is likely lower than that for HIP 3535 on a single night simply because HIP 58345 is a magnitude brighter in $V$.

\item {\it Stability over 2+ yr.}  The bottom panel of \rffigl{standards} includes data sequences for HIP 3535 (77 spectra, MAD value 15.2 \ms) and HIP 58345 (62 spectra, MAD 9.8 \ms) together, as observations are dove-tailed throughout the year to provide an unbroken series of RV standard observations. It is evident that there are several stretches of time when RV offsets are found in the HIP 3535 data sequence. We have examined various quantities in an attempt to reveal the cause(s) of the offsets for individual measurements. It appears that the drifts in the HIP 3535 data are {\it not} caused solely by (a) varying S/N in the spectra (primarily because nearly all spectra, 62 out of 77, have S/N > 50), out of the 15 RVs where the S/N is below 50, only four deviate by more than 15 \ms\ from the mean, (b) airmass, and (c) temperature changes in the Coude room where CHIRON is housed. We suspect that the poorer precision is due to shifts in the final spectral resolution: values computed from the individual spectra indicate that when the resolution dips by more than $\sim$1\%, the final RV points are offset. Resolution offsets occur when the focus of the spectrum onto the CCD drifts slightly, and shows that it is critical to keep the lines consistently as narrow as possible on the chip. Also we find a correlation between the most deviated RVs from the mean ($>$15 \ms) and their individual uncertainties. This is consistent with the fact that our error bars reflect mostly photon noise in combination with instrumental errors, but having bright stars in this case, the latter reason seems to dominate.
\end{enumerate}

In summary, for K dwarfs with $V$ = 7--12, the data indicate that CHIRON is stable to 5--20 \ms\ over timescales of hours, one month, and more than two years.

\vskip38pt

\section{K Dwarf Survey and Observations}

\subsection{Sample}
\label{sec:Sample}
The observed stars are a portion of an effort targeting more than 5000 K dwarfs within 50 pc; details about the full sample will be given in a future paper in this series. The particular subset observed here includes 190 stars selected from {\it Hipparcos} \citep{vanleeuwen:2007} to have parallaxes of at least 30 mas, placing them within 33 pc of the Sun, and that are between $+$30 and $-$30 deg declination. This equatorial sample has been selected so that each star can be observed from major observatories in both hemispheres. Sample star observations began before {\it Gaia} Data Release 2 \citep{evans:2018} results were available; hence, stars closer than the sample horizon entered the observing list using {\it Hipparcos} parallaxes.

Stars of spectral type K were chosen using an assessment of the regions where dividing lines between the G/K and K/M spectral types are found, with types determined by \cite{gray:2009} and RECONS. To define the blue and red ends of the K dwarf sequence, stars with spectral types were matched to  members of the RECONS 25 pc sample that have been carefully vetted for close companions, enabling us to use presumably single stars uncorrupted by close companions to map K dwarf spectral types to $V-K$ colors. We find that K dwarfs span $V-K$ = 1.90--3.70, and apply an additional constraint of $M_V$ = 5.80--8.80 to eliminate evolved stars and white dwarfs. The $V-K$ values were accumulated for the sample stars using $V$ from Tycho \citep{hog:2000} and $K$ from 2MASS \citep{skrutskie:2006}. The resulting list of 300 stars includes 110 that have stellar companions (either published by others or to be published by us) and/or have at least 10 RV measurements found in the data archives of HARPS \citep{mayor:2003} and HIRES \citep{vogt:1994}. The remaining 190 stars comprise the sample discussed here and are (a) within 33 pc, (b) located in the equatorial region of the sky, (c) have colors and absolute magnitudes of K dwarfs, and (d) have not been observed extensively, if at all, in previous RV programs.

The sample is presented in \rftabl{sample}, including 186 K dwarfs for which no companion has been detected with RV and four stars with new jovian exoplanet candidates. In addition, we provide results for nine more stars:  one K dwarf from \tess\ discovered to have a planet,  five K dwarfs with previously known planet candidates used as checks on our observing and reduction methodologies, and the three K dwarfs used as RV standards discussed in $\S$3.4. For all 199 stars, \rftabl{sample} provides names, coordinates, $V$ photometry from Tycho \citep{hog:2000}, $K_s$ photometry from 2MASS \citep{skrutskie:2006}, $B_g$ and $R_g$ photometry from $Gaia$ \citep{evans:2018}, the number and time coverage of the RV observations, and the mean absolute deviation (MAD) values for the RV series.

\subsection{Observations and Radial Velocity Precision}
\label{sec:Observations and Radial Velocity Precision}

\begin{figure}[ht!]
    \centerline{\includegraphics[width=0.47\textwidth]{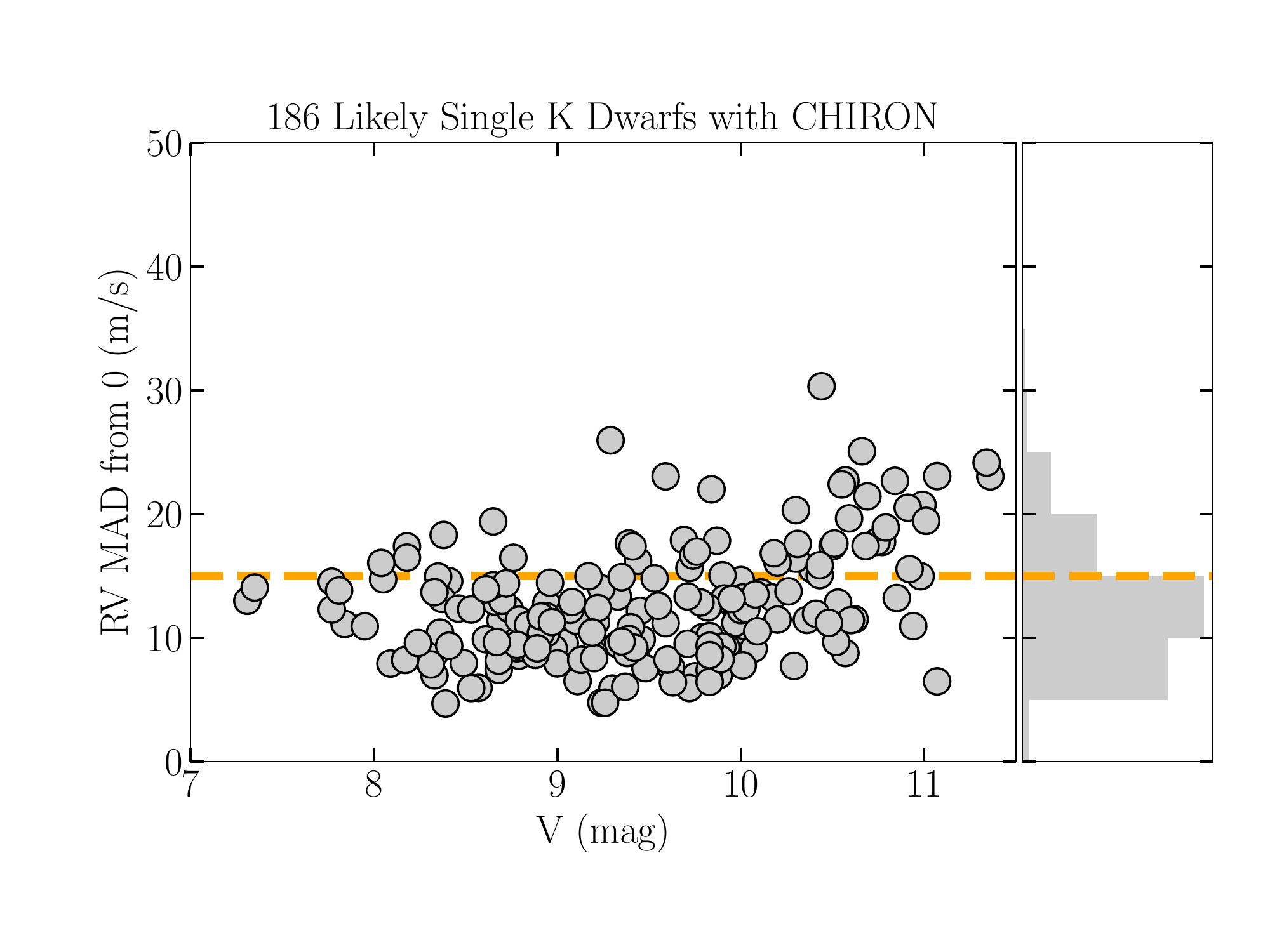}}
    \label{fig:rvmadv}
    \caption{Dependence of the scatter in the RV time-series with the $V$ magnitude for each of the 186 K dwarfs with no Keplerian RV signal (Table \ref{tab:sample}). The right panel groups data points in bins of 5 \ms, indicating that 75\% of radial velocity mean absolute deviations (MAD) are below 15 \ms\ (dashed line), with the best series as low as 5 \ms.}
\end{figure}

Our RV search is designed to perform a systematic reconnaissance for companions, so each K dwarf in our sample gets at least 6, and ideally 9, observations using the slicer instrument setup described in \rfsecl{setup4survey}. The observing cadence goal is to secure 2--3 spectra within 7 days, then repeat the sequence after a month, and then repeat it again after a year. The MAD values for the sequences of spectra for 186 stars for which we do not find periodic RV variations with CHIRON are illustrated in \rffigl{rvmadv} and listed on \rftabl{sample}. Given that jupiter-mass planets in (edge-on) orbits with periods of 10--100 days around K dwarfs cause RV variations of $\sim$40--100 \ms\, it is clear that CHIRON is precise enough to detect jupiter-mass exoplanets orbiting virtually all of the targeted K dwarfs.

For observational planning, it is useful to examine CHIRON's precision as a function of target brightness. \rffigl{rvmadv} illustrates the dependence of the scatter for the RV time-series using the MAD values vs. $V$ magnitudes of the K dwarfs whose RV curve is flat, i.e., they do not show periodic variation nor trend in their RVs over time. For our observing and data reduction protocols, we reach precisions of 5--15 \ms\ for $V$ = 7.0--10.5, with slightly poorer precision from $V$ = 10.5--11.5, therefore, the points above $\sim$20 \ms\ in the plot may, in fact, be as yet unidentified perturbations or young active stars, and they may be worthy of a closer look; additional observations are planned for these stars.

\begin{figure}[htb!]
    \centerline{\includegraphics[width=0.47\textwidth]{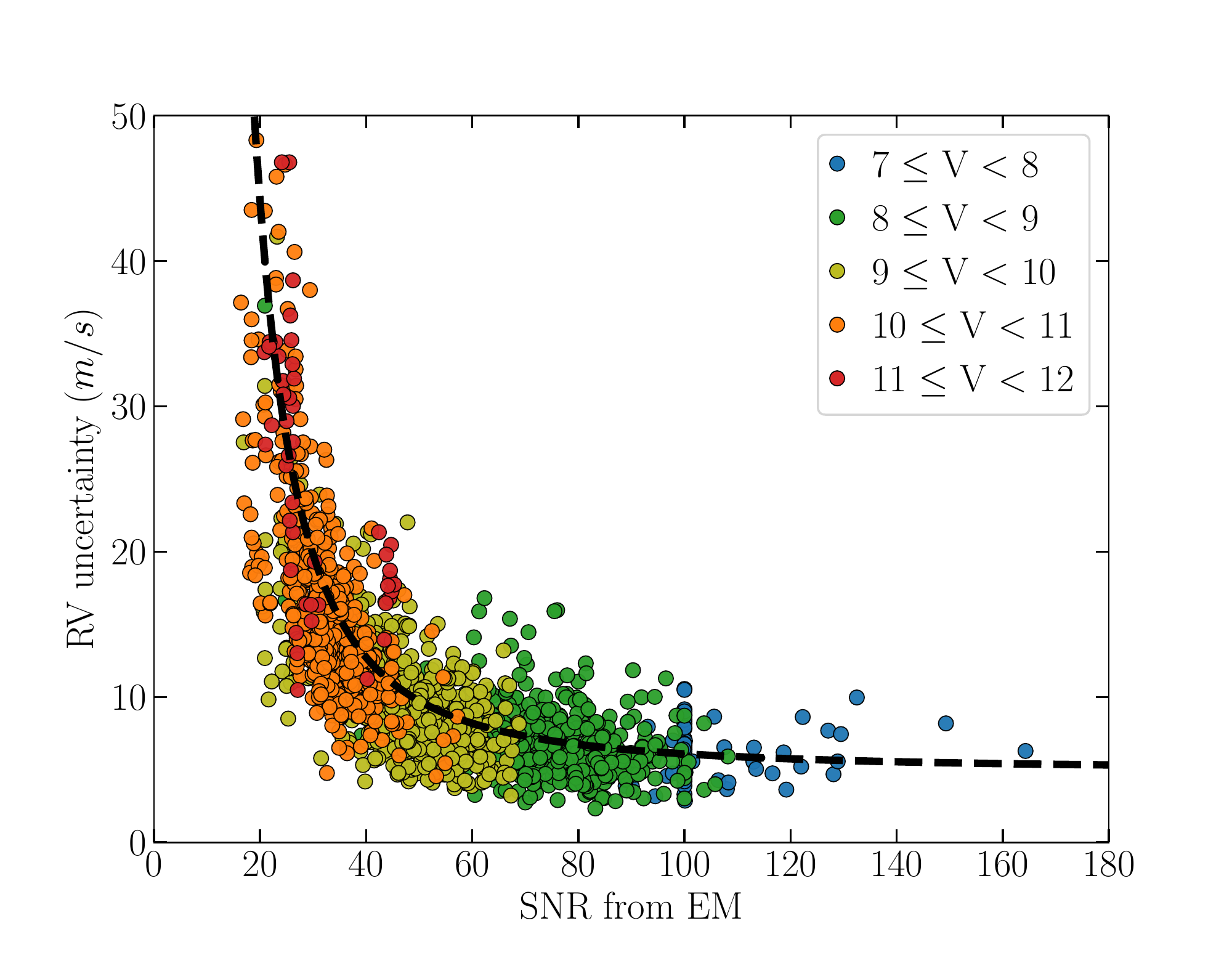}}
    \label{fig:rvunc_snr}
    \caption{S/N calculated from CHIRON's exposure meter (EM) versus RV uncertainty for individual spectra, color coded by stellar V magnitude. Each of the 1784 data points represents a single spectrum. The dashed line is described by Eq. \eqref{eq:rvunc_snr} and is derived using observations with exposure times of 900 seconds. The slight overabundance of points at S/N = 100 is the result of observations that were stopped before 900 seconds, when the S/N reached 100.}
\end{figure}

In addition to mapping the dependence of RV precision with target brightness, because sky conditions and telescope tracking vary, it is useful to map the precision as a function of the S/N for individual spectra (as measured by CHIRON's exposure meter during an observation). \rffigl{rvunc_snr} relates the S/N values for 1784 individual spectra of the 186 K dwarfs to the resulting RV uncertainties. Overall, for S/N values of at least 40, the uncertainties are less than 15 \ms, whereas for S/N $\sim$20, the uncertainties increase to 30 \ms\ and above. It is therefore recommended that to reach a precision of 15 \ms\ in slicer mode with CHIRON, observers targeting K dwarfs or similar stars anticipate exposure times of 900 sec (our standard exposure time) for stars with $V$ $\lesssim$ 10.5. As a rule of thumb, to obtain a single measurement error in RV of $\sim$5 \ms, it is necessary to reach S/N $\sim$100 at 5500 \AA. This is possible for a K dwarf brighter than V $\sim$ 9 in 900 seconds exposure in slicer mode. Overall, for stars like those observed in our program, the expected precision can be predicted using:

\begin{equation}
\label{eq:rvunc_snr}
{\sigma}_{RV} \sim \frac{10 000}{\textit{S/N}^2}+4 \ \ms
\end{equation}

The S/N of each spectrum is computed using the counts from the exposure meter, which picks off about 1\% of the collimated light at 5450\AA~with a bandwidth of 900\AA\ \citep{tokovinin:2013}. This S/N value maps directly to the S/N of the raw spectra and provides a straightforward method to evaluate individual exposures. Importantly, the S/N can be tracked by an observer during an exposure to compensate for cloud cover, seeing, and telescope tracking, all of which can affect how much light is injected into the fiber. As a result, an observer can actively stop or extend an integration to reach a desired S/N. For example, the collection of vertical points at S/N = 100 represents observations of relatively bright stars for which exposures were stopped before 900 seconds. The relation to convert from exposure meter counts to S/N is given by:

\begin{equation}
\textit{S/N} = \sqrt{\frac{\textit{EMNUMSMP} \times \textit{EMAVG} - 7401.973}{57.909}}
\end{equation}
\vskip10pt

where \textit{EMNUMSMP} is the number of samples of a tenth of a second during the exposure and \textit{EMAVG} is average instant number of photon counts during the exposure; both are recorded as header entries with these names for each spectrum taken. 

As outlined in \rfsecl{rvpipeline}, the individual RV measurements have been determined using the weighted standard error of the RVs measured for the 14 orders adopted in the pipeline. While there are other errors, such as systematic instrumental offsets to be considered in future work when larger datasets are available, for this characterization of our K dwarf survey prospects, we report errors that are only the result of the statistical results based on individual RVs extracted from the 14 different orders. As expected, the RV errors are lower for brighter stars and higher S/N spectra. 

\begin{figure}[ht!]
    \centerline{\includegraphics[width=0.47\textwidth]{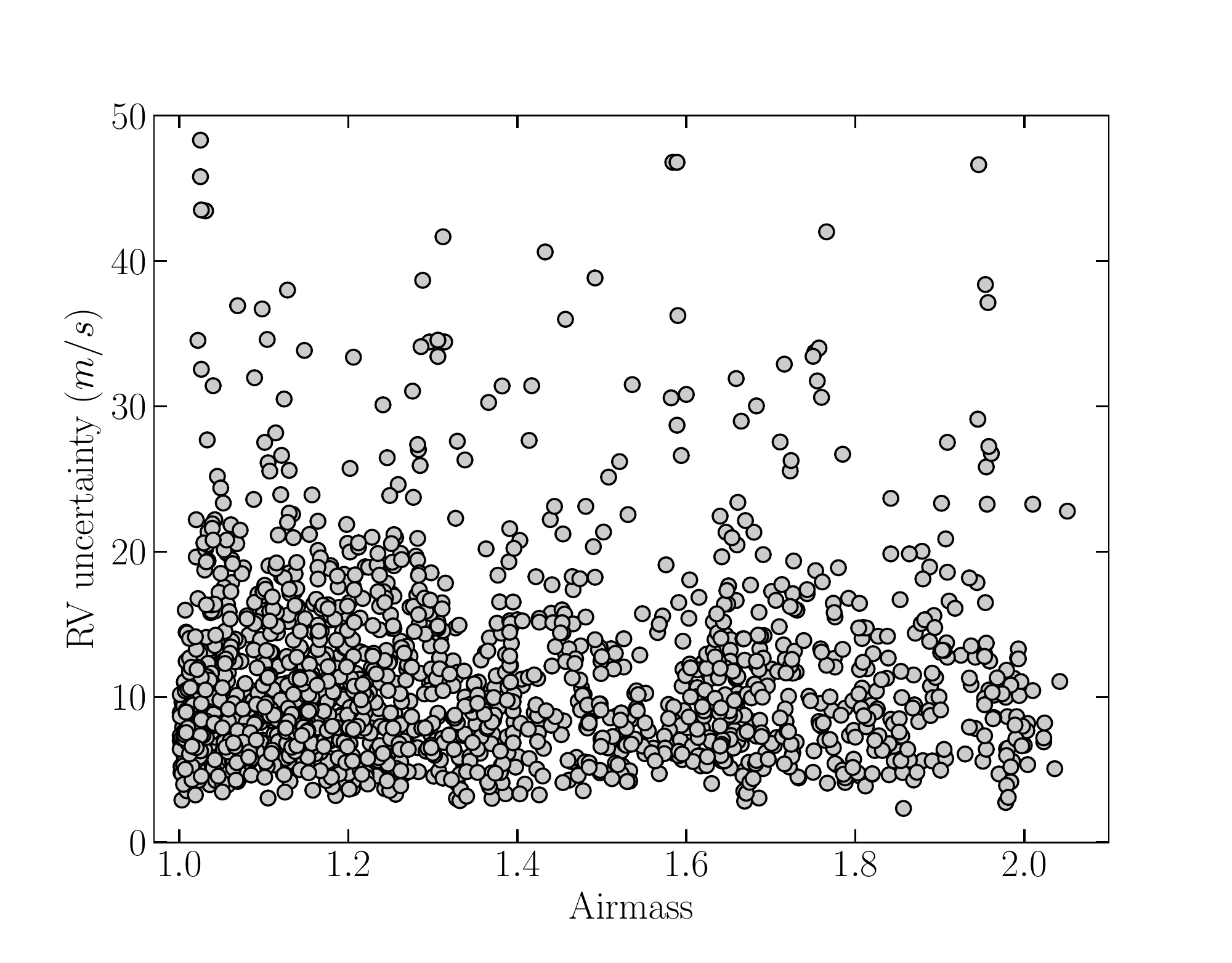}}
    \label{fig:rvunc_airmass}
    \caption{Airmass versus RV uncertainty per spectrum. Typically, targets are observed when they are close to the meridian passage, but is not always possible due to scheduling constraints. It is clear that RV uncertainty is not systematically affected by the airmass at which the K dwarfs have been observed.}
\end{figure}

Finally, because our sample stars span DEC = $-$30 to $+$30 deg, we observe stars with airmasses of 1.0--2.0 from CTIO. Although not every observation is timed precisely when a star passes through the meridian, most observations occur near transit, and a plot of airmass vs. RV uncertainty is useful to understand whether or not high airmass observations yield lower RV precision. \rffigl{rvunc_airmass} illustrates that the RV uncertainty is independent of the airmass, a result that is particularly encouraging because stars north of DEC = 0 can be observed with CHIRON as effectively as more southern stars. This also bodes well for individual targets for which many observations are desired over the span of a single night, given that RV uncertainties are consistent over a large range of airmass.

\section{Results}

Over the past two decades, more than 4300 planets orbiting other stars have been detected and confirmed (as listed in the NASA Exoplanet Archive, \href{exoplanetarchive.ipac.caltech.edu}{exoplanetarchive.ipac.caltech.edu} as of May 2021), including more than 3300 revealed via the exoplanet transit method. The greatest contributor of transiting exoplanets has been the \kep\ mission \citep{borucki:2010}, which was extended via the \ktwo\ effort \citep{howell:2014}. Together, \kep\ and \ktwo\ have  revealed thousands of exoplanet candidates, and \tess\ continues to add to the candidate list. Among the finds are terrestrial, ice giant, and gas giant planets, with orbital periods ranging from hours to years. Most relevant to the work presented here, given the precision of CHIRON and the duration of the survey, are planet candidates with masses of $0.3\ \mjup < m\sin{i} < 10\ \mjup$ and orbital periods of $P < 180$ days. As of May 2021, a search of the NASA Exoplanet Archive yields 93 stellar hosts within 100 pc with at least one confirmed exoplanet that meet these criteria.

\begin{figure*}[ht!]
\centerline{
\includegraphics[width=0.43\linewidth]{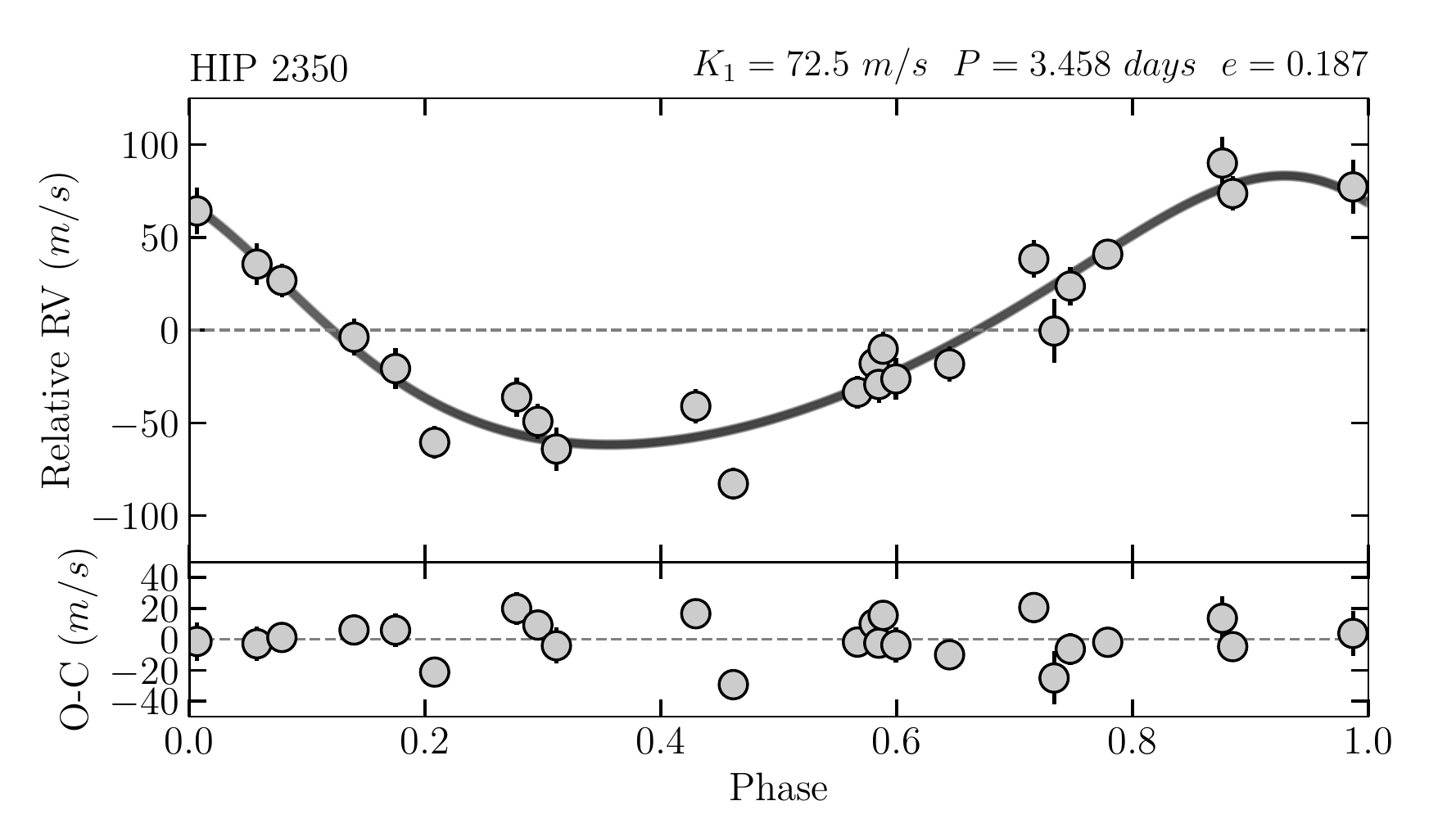}
\includegraphics[width=0.43\linewidth]{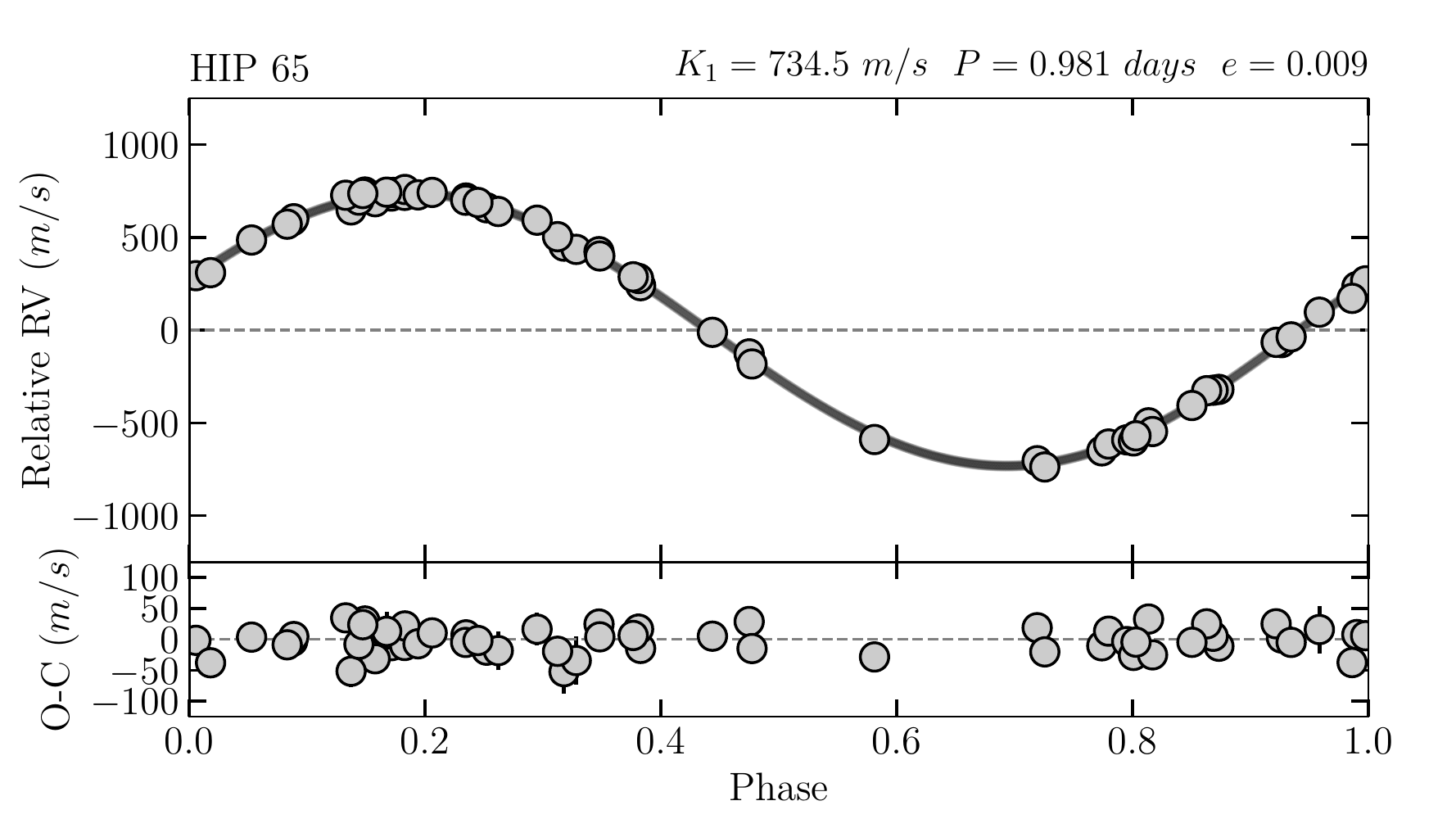}}
\centerline{
\includegraphics[width=0.43\linewidth]{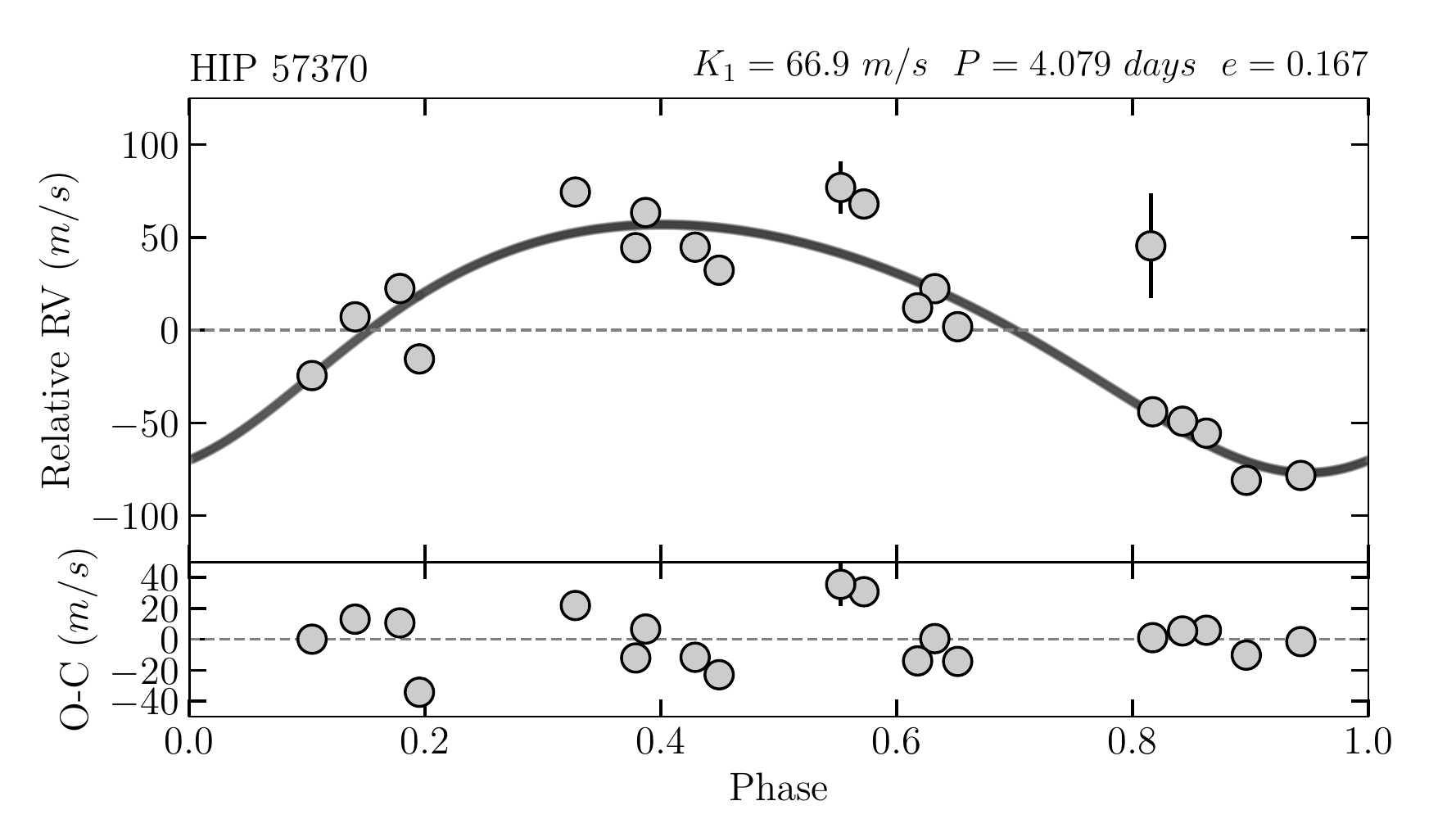}
\includegraphics[width=0.43\linewidth]{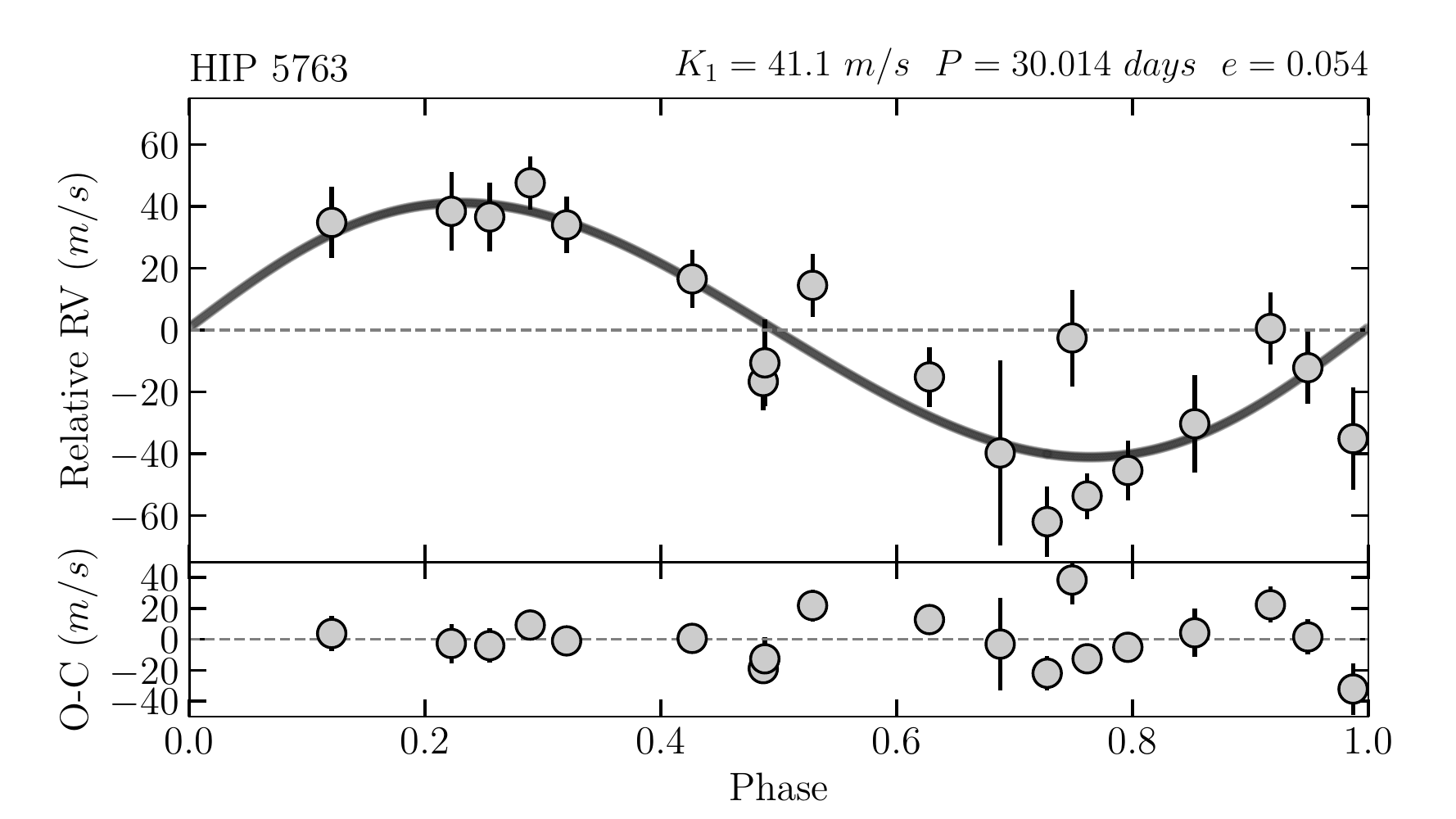}}
\centerline{
\includegraphics[width=0.43\linewidth]{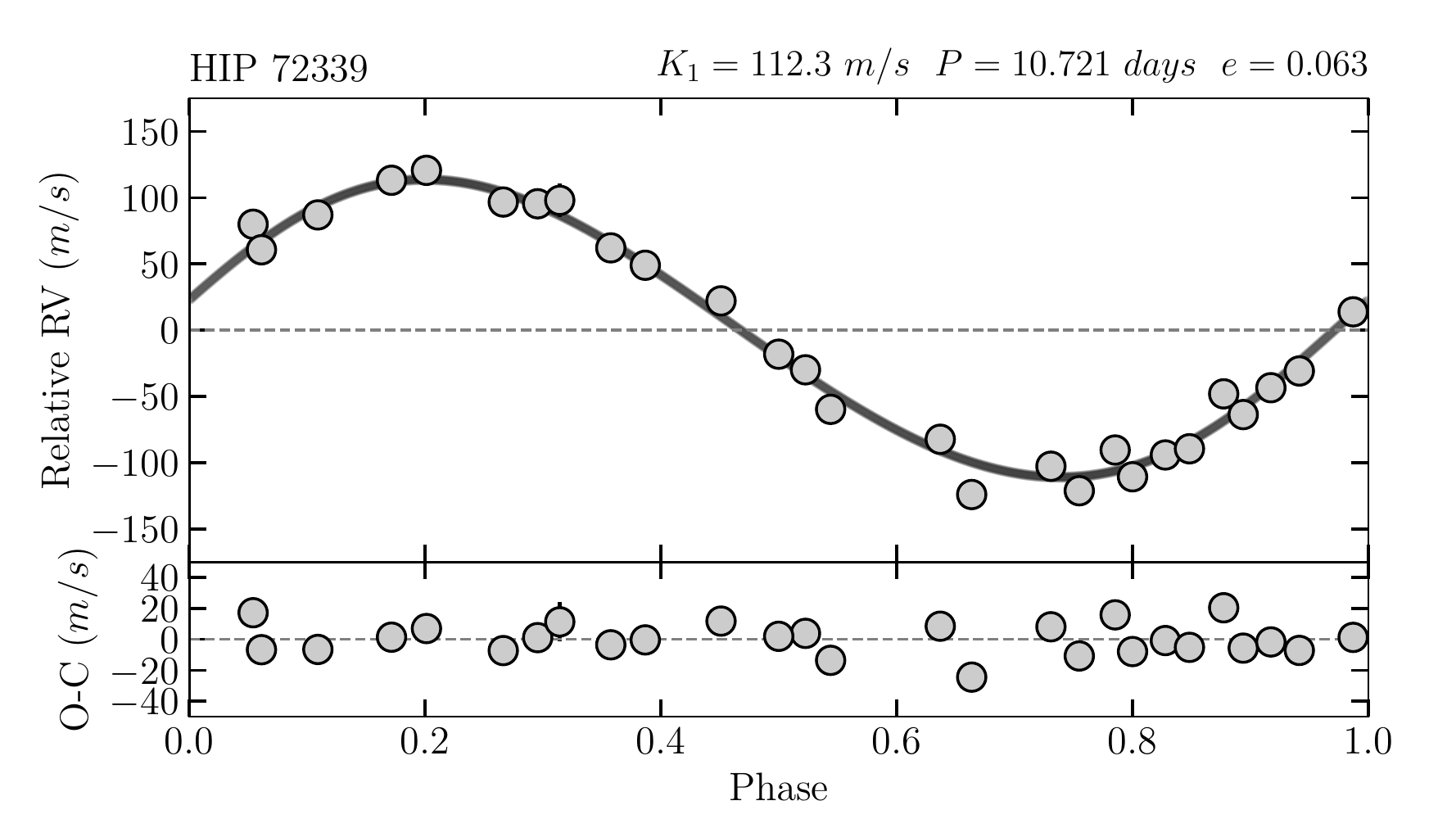}
\includegraphics[width=0.43\linewidth]{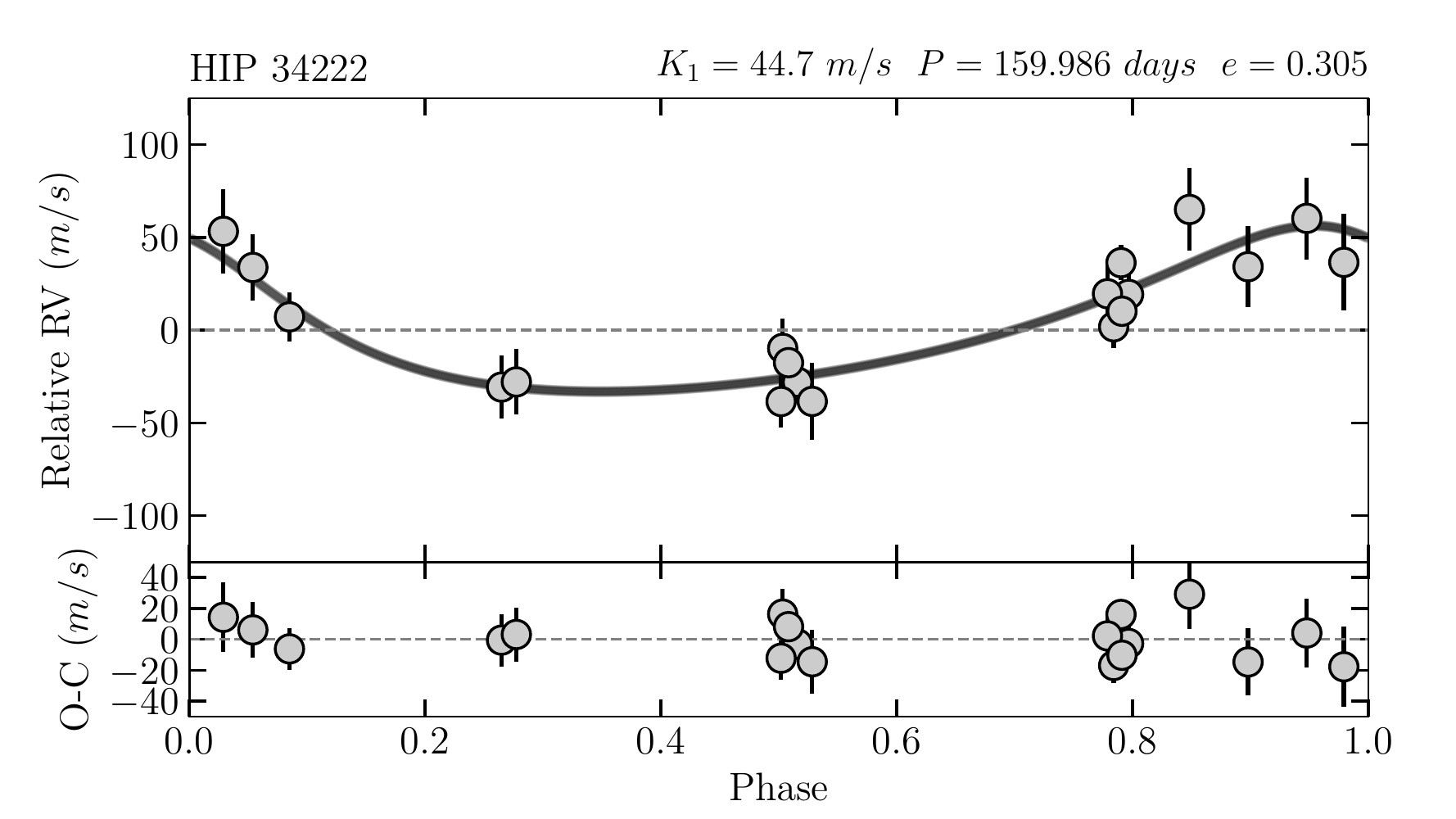}}
\centerline{
\includegraphics[width=0.43\linewidth]{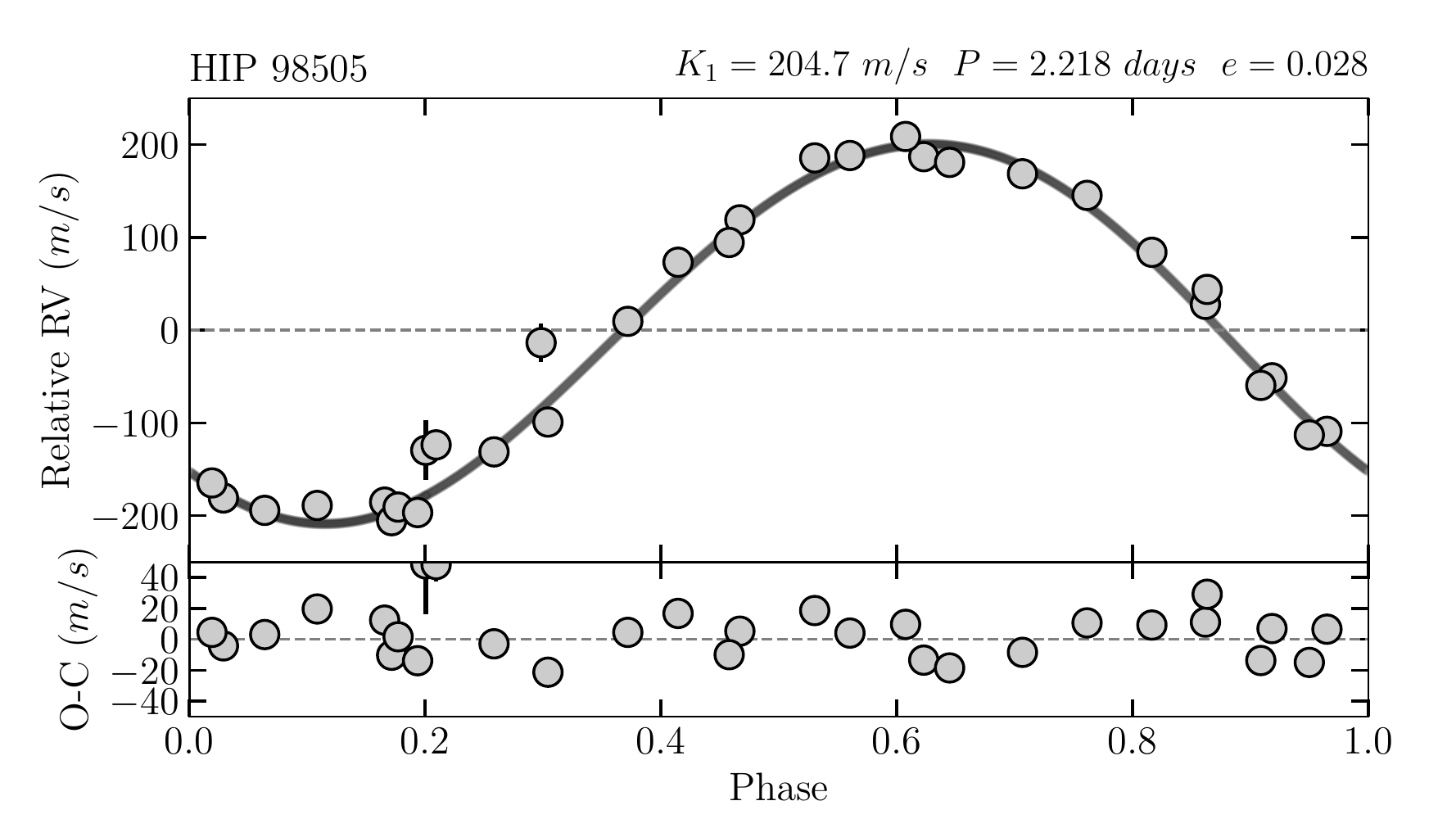}
\includegraphics[width=0.43\linewidth]{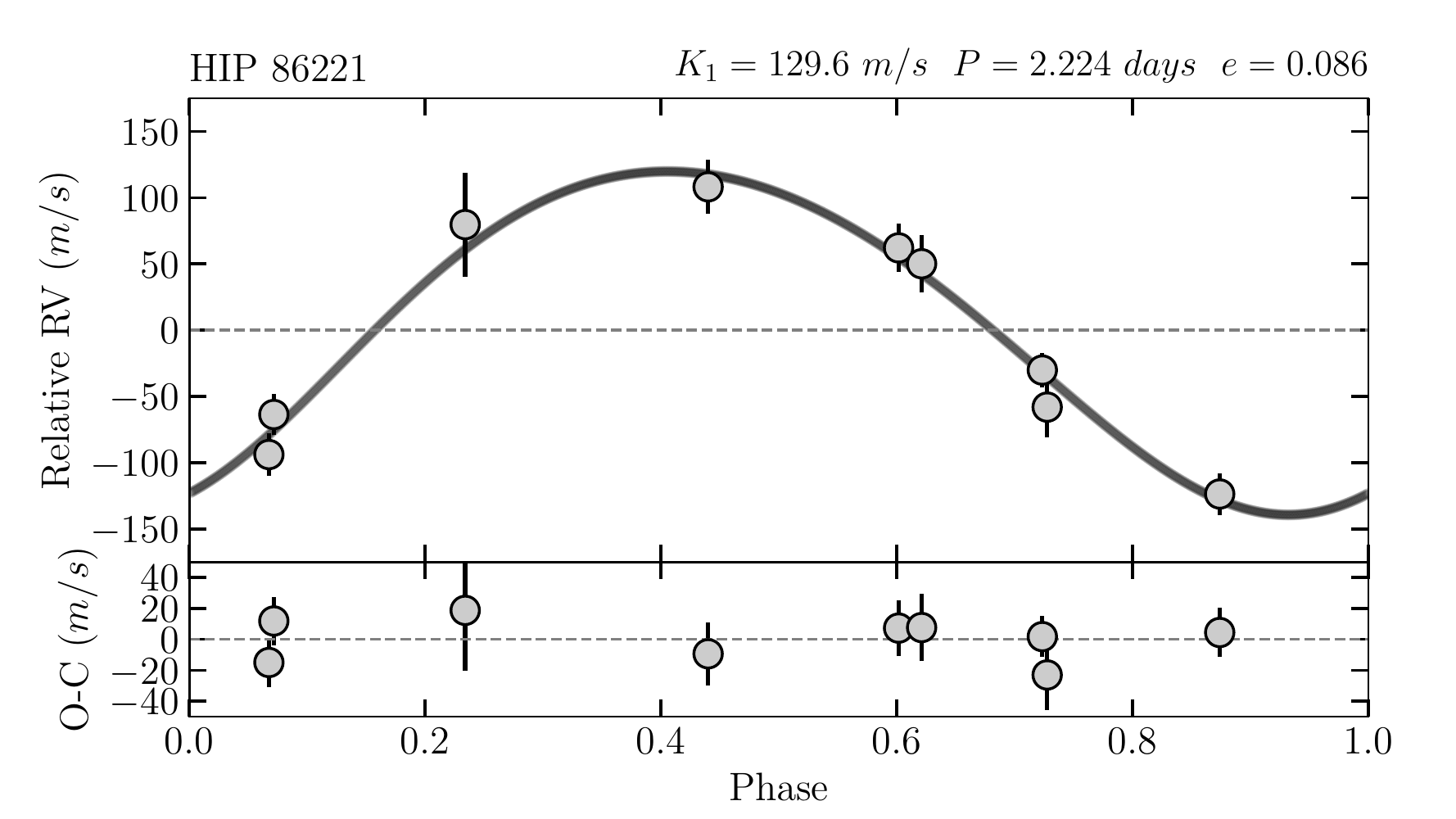}}
\centerline{
\includegraphics[width=0.43\linewidth]{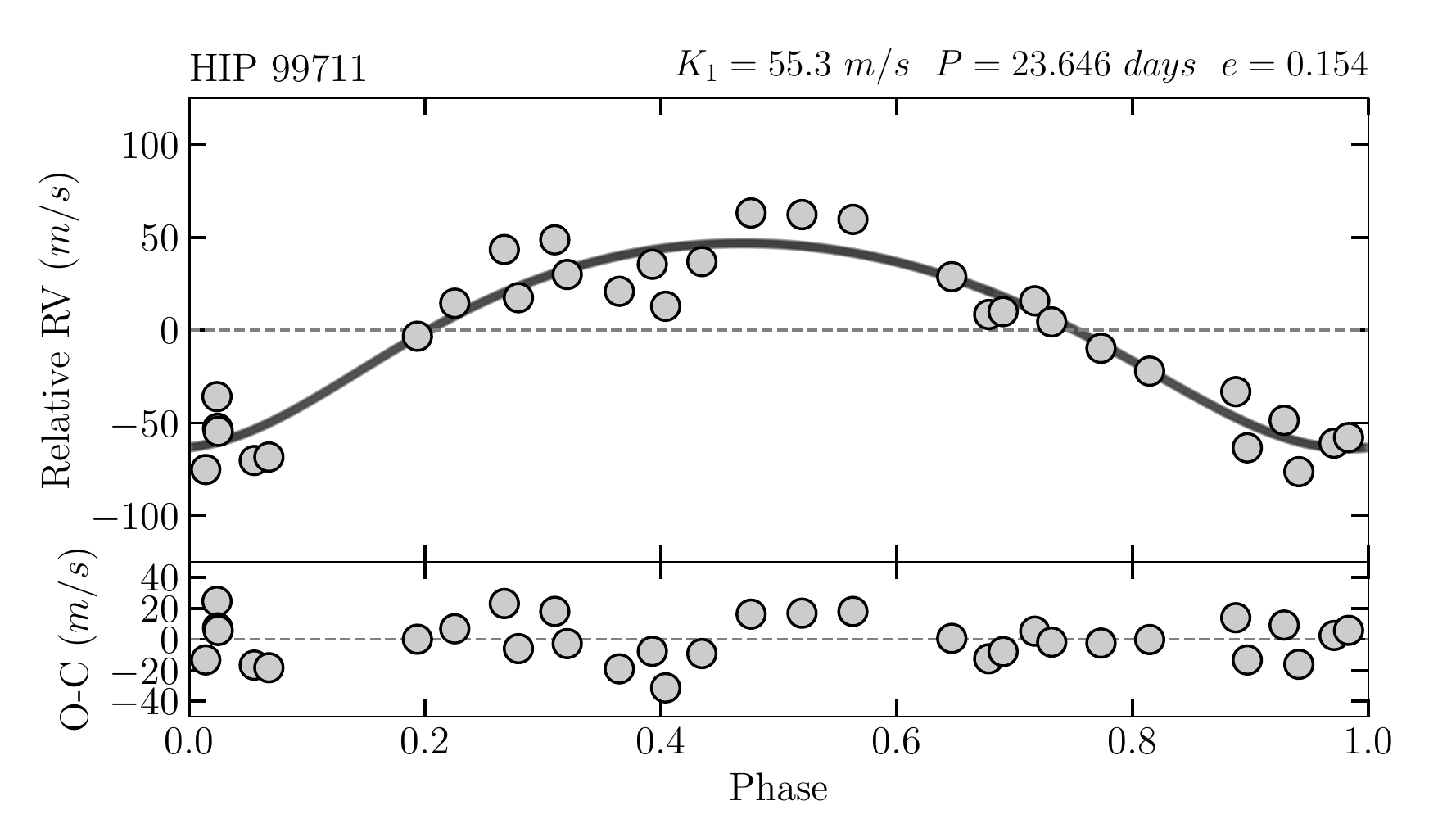}
\hspace{0.43\linewidth}}
\label{fig:orbfits}
\caption{Phase-folded RV curves and residuals derived from CHIRON spectra. Phase zero indicates the time of the periastron passage ($T_p$). \textit{Left column}: five planets known to orbit K dwarfs. \textit{Right column}: four new planet candidates around K dwarfs.}
\end{figure*}


Many of the exoplanets revealed during the transit surveys are followed up using RVs to confirm that the object is, indeed, a planet orbiting the star identified to exhibit the transit. In addition,  there have been more than 800 candidates detected during RV surveys that do not transit. Here we discuss nine K dwarfs that are orbited by low mass companions that are exoplanet candidates. A total of 240 individual measured RVs from CHIRON are given for these nine stars in \rftabl{rvs}, ordered by $Hipparcos$ number. 

\begin{deluxetable*}{cccccccccccc}[ht!]
\setlength{\tabcolsep}{0.05in}
\tabletypesize{\scriptsize}
\tablecaption{\label{tab:orbpars} Orbital Parameters of Exoplanet Candidates.}
\tablehead{\colhead{Star} &
           \colhead{$M_*$} &
           \colhead{Period} & 
           \colhead{$m \sin{i}$} & 
           \colhead{$e$} & 
           \colhead{$\omega$} & 
           \colhead{$K$} & 
           \colhead{$a$} & 
           \colhead{$T_p$} & 
           \colhead{RMS} & 
           \colhead{Data points} & 
           \colhead{Ref.} \\ 
           \colhead{} &
           \colhead{$M_\odot$} &
           \colhead{(days)} & 
           \colhead{(\mjup)} & 
           \colhead{} & 
           \colhead{(deg)} & 
           \colhead{(\ms)} & 
           \colhead{(AU)} & 
           \colhead{(JD - 2450000)} & 
           \colhead{(\ms)} & 
           \colhead{} & 
           \colhead{}} 

\startdata
\hline
\multicolumn{12}{c}{Previously Known Exoplanet Systems}
\\
\hline
HIP 2350  &  0.92  &   3.458$_{-0.0083}^{+0.0081}$  & 0.50  &  0.187   &  37.28  &  72.5$_{ -4.4}^{ +4.6}$  & 0.044  & 7948.482 & 12.8 &  24 & this work                   \\ %
          &  0.93  &   3.444                      & 0.48  &  0.000   & 126.90  &  67.4                  & 0.044  & 3323.206 & ...  &  28 & \cite{moutou:2005}  \\ %
HIP 57370 &  0.92  &   4.079$_{-0.0146}^{+0.0142}$  & 0.49  &  0.167   & 206.16  &  66.9$_{ -3.3}^{ +3.5}$  & 0.049  & 7933.659 & 26.1 &  20 & this work                   \\ %
          &  0.93  &   4.114                      & 0.49  & <0.140   & 143.40  &  63.4                  & 0.049  & 3732.700 & 16.0 &  59 & \cite{ge:2006}      \\ %
HIP 72339 &  0.85  &  10.721$_{-0.0031}^{+0.0033}$  & 1.09  &  0.063   & 281.05  & 112.3$_{ -4.5}^{ +4.6}$  & 0.090  & 7928.965 &  9.9 &  28 & this work                   \\ %
          &  0.79  &  10.720                      & 1.02  &  0.044   & 203.63  & 115.0                  & 0.088  & 1287.380 & 15.4 & 118 & \cite{udry:2000}    \\ %
HIP 98505 &  0.84  &   2.218$_{-0.0009}^{+0.0010}$  & 1.17  &  0.028   & 136.42  & 204.7$_{ -2.5}^{ +2.6}$  & 0.031  & 7929.288 & 21.5 &  31 & this work                   \\ %
          &  0.82  &   2.219                      & 1.15  &  0 (fix) &    ...  & 205.0                  & 0.031  & ...      & ...  &  35 & \cite{bouchy:2005}  \\ %
HIP 99711 &  0.74  &  23.646$_{-0.2082}^{+0.2304}$  & 0.63  &  0.154   & 188.21  &  55.3$_{ -2.2}^{ +2.1}$  & 0.146  & 7912.820 & 13.5 &  32 & this work                   \\ %
          &  0.75  &  24.348                      & 0.72  &  0 (fix) &   0.00  &  61.0                  & 0.150  & ...      & ...  & 182 & \cite{santos:2003}  \\ %
\hline
\multicolumn{12}{c}{TESS Exoplanet System}
\\
\hline
HIP 65    &  0.74  &   0.981 (fix)                & 2.95  &  0.009   & 291.42  & 734.6$_{ -4.5}^{ +4.6}$  & 0.017  & 8368.833 & 20.4 &  58 & this work                   \\ %
          &  0.78  &   0.981                      & 3.21  &  0 (fix) &    ...  & 753.7                  & 0.017  & ...      & ...  &  34 & \cite{nielsen:2020} \\ %
\hline
\multicolumn{12}{c}{New Candidate Exoplanet Systems}
\\
\hline
HIP 5763  &  0.72  &  30.014$_{-0.2842}^{+0.1528}$  & 0.51  &  0.054   & 271.08  &  41.1$_{-11.2}^{ +8.8}$  & 0.170  & 8056.731 & 16.2 &  19 & this work                   \\ %
HIP 34222 &  0.62  & 159.986$_{-2.9256}^{+2.6753}$  & 0.83  &  0.305   &  31.80  &  44.7$_{ -8.3}^{ +9.1}$  & 0.492  & 8023.220 & 12.7 &  19 & this work                   \\ %
HIP 86221 &  0.79  &   2.224$_{-0.0005}^{+0.0004}$  & 0.71  &  0.086   & 208.93  & 129.6$_{-19.6}^{+34.5}$  & 0.031  & 7947.283 & 12.8 &   9 & this work                   \\ %
\enddata
\end{deluxetable*}


The orbital fits for all nine systems are shown in \rffigl{orbfits}. Our orbits have been derived using the code Systemic2 \citep{meschiari:2009}\footnote{\href{https://github.com/stefano-meschiari/Systemic2}{github.com/stefano-meschiari/Systemic2}}, which provides a functional interface that can be used to calculate Lomb-Scargle periodograms and to explore orbital fits interactively given a set of RV data. All plots have zero phase defined to be at the epoch of periastron. Our calculated RVs are input into the code and the periodogram is inspected for prominent peaks above the 10\% false alarm probability (FAP) level. We avoid those peaks that match the nightly cadence and total baseline of the observations because they are caused by sampling aliases. Keplerian orbits are then fit for each of $\sim$5 strongest peaks and the results inspected by eye. Initial fits are made using circular orbits in order to avoid very high eccentricity orbits that may fit the datasets but are astrophysically unlikely. Step sizes of 0.001 days are used to fine tune the orbital period, while the fitting process is carried out using chi-square minimization until each orbit fit converges, determined when the RMS of the fit reaches $\sim$20 \ms. We have chosen this limit given the typical RV scatter and uncertainties seen for the 186 K dwarfs with no detected companions, as shown in \rffigl{rvmadv} and \rffigl{rvunc_snr}. As a final step in the orbital parameter determinations, we derive the intervals of confidence for parameters using MCMC simulations also included in Systemic2 code, starting with the model best fitted, a total $\sim$10000 steps, in 2 chains, and skipping the first 1000 iterations. Errors are not listed for previous orbital parameters because they are a mix of different types of errors.

\rftabl{orbpars} includes the orbital parameters and errors for all nine K dwarf + exoplanet systems highlighted here. To estimate the companion $m \sin i$ values, masses for the primary stars have been derived using $V$ magnitudes converted from $V_T$ magnitudes \citep{hog:2000} and the mass-luminosity relation in $M_V$ of \cite{henry:1993}. A future paper is planned to provide a much-needed update to the mass-luminosity relation for K dwarfs using empirically determined masses.

\subsection{Known Planets Orbiting K Dwarfs}

Here we provide five examples of previously known planet candidates orbiting K dwarfs revealed via RVs. All five are (presumed) single planet systems chosen during the initial reopening of the CTIO/SMARTS 1.5m in 2017 June to serve as test targets to verify the efficacy of CHIRON under typical observing protocols. The instrument setup, data reduction, and RV calculation methods were as described in sections \rfsecl{specsandsetup} through \rfsecl{rvpipeline} for all of the stars. Each of the new orbital solutions presented here has been calculated solely based on the CHIRON observations and therefore are independent from previously reported solutions found in the literature.

\subsubsection{HIP 2350}

This star ($V$ = 9.37, K1V) was reported by \cite{moutou:2005} to have a hot Jupiter of minimum mass 0.48 \mjup\ with an orbital period of 3.444 days. We obtained 24 spectra of HIP 2350 between 2017 July 16 and August 08 and confirm a 0.50 \mjup\ minimum mass planet with an orbital period of 3.458 days, consistent with the previous result. The 73 \ms\ RV semi-amplitude is $\sim$7 times larger than the typical uncertainty for K dwarfs in our program, indicating that hot Jupiters of this type are easily revealed by CHIRON. HIP 2350 has a lower mass stellar companion detected with RoboAO at Palomar \citep{baranec:2012} at a separation of 0.5\arcsec\ and fainter by 3.3 mag at 754 nm (Sloan i' filter) by \cite{riddle:2015} and \cite{roberts:2015}. The companion was confirmed with DSSI at Gemini North \citep{horch:2009} at a separation of 0.5\arcsec\ and fainter by 3.8 mag at 692 nm \citep{wittrock:2016}. The companion is likely an M1V with an estimated orbital period of $\sim$130 yrs from the projected separation, and it should not pose a serious threat to the detection, nor the orbital stability of the planet. \cite{shaya:2011} and later \cite{oh:2017} reported HIP 2350 to be part of a co-moving wide pair where the primary component is HIP 2292, another solar-type star, at projected separation of 897\arcsec\ ($\mysim$0.2 pc). While the system could still be a wide multiple, \cite{oh:2017} also discuss the possibility that systems such as these could have formed together but are now drifting apart.

\subsubsection{HIP 57370}

This star ($V$ = 8.05, K0V) was reported by \cite{ge:2006} to have a hot Jupiter with a minimum mass 0.49 \mjup\ in an orbit with a period of 4.114 days. We obtained 20 spectra of HIP 57370 between 2017 June 29 and 2017 July 26, and the same minimum mass of 0.49 \mjup\ in an orbital period of 4.079 days. The single point with a large error near phase 0.8 in the panel for this star in \rffigl{orbfits} was taken at the beginning of the night on 2017 July 26, when clouds were present and the seeing expanded to 2.5\arcsec, resulting in poor S/N for this observation. This data point is included to illustrate the relative quality of a poor observation suffering from high photon noise error compared to more typical measurements. The lack of speckle companions (Nusdeo et al.~2020, submitted), nor any known visual stellar companions makes this detection robust. Despite being a non-transiting planet, \cite{guilluy:2019} uses this hot Jupiter (HD 102195 b) to demonstrate the feasibility of detailed studies of exoplanet atmospheres using the GIANO spectrograph \citep{oliva:2006} mounted at Telescopio Nazionale Galileo (TNG), a 4-m class telescope.

\subsubsection{HIP 72339}

This star ($V$ = 8.04, K0V) is near the celestial equator, but at $V-K_s$ = 1.81 is slightly bluer than the blue cutoff we used for our K dwarf sample; thus, it is not part of the larger survey but was observed strictly as a benchmark. This star has a known hot Jupiter with minimum mass 1.02 \mjup\ and orbital period 10.720 days \citep{udry:2000}. We confirm the companion via 28 spectra to have a somewhat larger minimum mass of 1.02 \mjup\ and a virtually identical orbital period of 10.721 days. Our phased RV curve spans 3.5 full orbits for data taken in 2017 from June 24 to August 7, and this is the star with a known planet for which we have the longest time coverage at 42 days. Given the RV variation caused by the companion of $K$ = 112 \ms, this Jupiter-like is clearly detected, indicating that CHIRON reveals such candidates easily. No visual stellar companions are known in the system and the planet is not found to be transiting. While this system has been revisited by \cite{wittenmyer:2009} and \cite{hinkel:2015}, the candidate exoplanet's properties have not changed significantly since the initial discovery.

\subsubsection{HIP 98505}

We obtained 31 observations of HIP 98505 ($V$ = 7.66, K2V) between 2017 June 24 and August 4, spanning 40 days, to reveal a companion with minimum mass 1.17 \mjup\ in a 2.218 day orbit. This planet (HD 189733 b) was discovered by \cite{bouchy:2005} and reported to have a minimum mass of 1.15 \mjup\ in a 2.219 day orbit; the system has been extensively studied since then with no significant changes in the orbital parameters. At an orbital inclination of $\mysim$85 degrees, the exoplanet transits its host star, allowing detailed determination of its fundamental parameters, which combined with the proximity and brightness of the star have made the system an ideal laboratory for exoplanet atmosphere studies (e.g., \citealt{redfield:2008} and \citealt{guilluy:2020} and references therein). \cite{bakos:2006} reports a lower mass companion 11.2\arcsec\ from HIP 98505 that is 3.7 magnitudes fainter in $K_s$. Using astrometry, proper motion, radial velocity, and photometry, they derive an orbit for the stellar companion nearly perpendicular to the planet's transiting orbit, i.e. nearly face-on. However, at a projected separation of 218 AU and orbital period of 3200 years, the orbit is highly uncertain.

\subsubsection{HIP 99711}

We found for this star ($V$ = 7.76, K2V) a 0.63 \mjup\ companion with the longest orbital period (23.646 days) and the smallest RV amplitude (55.3 \ms) of the five selected benchmark stars. Nonetheless, the RV perturbation is clear in the CHIRON data, and the values are close enough matches to those in the discovery paper \citep{santos:2000}. The companion (HD 192263 b) is among the earliest exoplanet candidates, but was called into question by \cite{henry:2002}, who argued that the RV signal detected is due to stellar magnetic activity rather than the stellar reflex motion caused by a companion. However, \cite{santos:2003} then provided further proof to improve the planet's orbit, and confirmed the discovery, which is also supported by our measurements. We suspect that the difference in the orbital solution we found is due to contamination of the RV signal by the same stellar magnetic activity that initially disputed the discovery.

\subsection{New Planets Orbiting K Dwarfs}
Here we discuss four candidates for giant planets orbiting nearby K dwarfs, including a contemporaneous detection of a transiting exoplanet from \tess\ and three new candidate exoplanets from our survey.

\subsubsection{\tess\ target HIP 65 (NLTT 57844)}

A possible low mass companion was found to transit HIP 65 (TOI 129, $V$ = 11.13, K4V) in the first set of data released from \tess\ in sector 2, and later in sectors 28, 29. At 61.9 pc, this star is beyond the 50 pc cutoff of our K dwarf survey, but was observed as an early possible discovery by \tess\ that could be quickly verified with CHIRON data. Initial RVs were collected with CHIRON starting 2018 September 8 and within two weeks of the first \tess\ data release, an orbit was published at \href{www.recons.org}{www.recons.org} on 2018 September 10. The quick turnaround was possible because of the nimble system established to acquire CHIRON observations. A total of 58 spectroscopic observations have been secured between 2018 September 8 to 2020 January 12 and we find an RV signal consistent with the \tess\ transit signal detected. Our analysis yields to a giant planet companion with mass 2.95 \mjup~in an orbital period of 0.981 days, consistent with the orbit published by \cite{nielsen:2020}. The properties of the $\sim$1 day orbit and massive planet make this an excellent candidate for detailed exoplanet atmospheric studies.

\subsubsection{HIP 5763}

This K dwarf survey star ($V$ = 9.86, K6V) shows a perturbation with a period 30 days in the RVs due to a companion with minimum mass 0.51 \mjup. A total of 19 observations spanning two years were secured between 2017 November 20 and 2019 December 17. This planet candidate has the smallest RV amplitude (41.1 \ms) of the four new detections reported here, but this amplitude is clearly offset from results for stars of similar brightness in \rffigl{rvmadv}, indicating that the companion is likely real. In addition, the RMS of the residuals to the orbital fit (16.2 \ms) is similar to the RMS values we find after fitting orbits for the five known planetary systems. HIP 5763 is not known to have visual stellar companions reported in the Washington Double Star Catalog (WDS) \citep{mason:2001}, nor any spectroscopic binary companions reported in The Ninth Catalogue of Spectroscopic Binary Orbits (SB9) \citep{pourbaix:2004}. \tess\ observed this target in sector 17 from 2019 October 8 to 2019 November 2 (25 days) at 2 minute cadence; the \tess\ SPOC (Science Processing Operations Center) pipeline does not report transit events.

\subsubsection{HIP 34222}

At $V$ = 10.23, this K7V star is one of the fainter K dwarfs in the sample, which is reflected in the relatively large error bars on individual points. We obtained 19 spectra spread over two years between 2017 December 15 and 2019 December 18. The dataset indicates a possible companion with minimum mass 0.83 \mjup~in an orbit with $e$ = 0.301, which is the most eccentric of the nine systems discussed here. With a derived orbital period of 160 days, the relatively high eccentricity is not precluded by tidal circularization, which happens only for systems with orbital periods less than a few weeks \citep{halbwachs:2005}; nonetheless, we consider this to be the least precise orbit presented here. The WDS reports WDS J07057+2728B as a visual companion 4.6 magnitudes fainter at 13.5\arcsec, but no parallax nor reliable proper motion is available to confirm it is bound to HIP 34222. Although WDS lists five stars as nearby, none are physical companions. No spectroscopic companion is listed in SB9 catalog.

\subsubsection{HIP 86221}

This star ($V$ = 9.20, K5V) is among the most northern in our equatorial sample, with DEC $\sim$ $+$28 deg. We find a classic hot Jupiter candidate in a 2.2 day orbit with minimum mass 0.71 \mjup. Although we have only 9 observations to date for this star, the semi-amplitude of the orbital fit, 130 \ms, is more than 10 times CHIRON's typical MAD value for K dwarfs of this brightness, so we consider the detection secure. The HIP 86221 system is known to be a stellar triple. The AB components are separated by a few tenths of an arcsecond, with B fainter than A by 0.59 mag in the optical. A visual binary orbit has been determined for AB using astrometry and speckle interferometry, yielding a period of 23.991 years, semi-major axis of 0\farcs2884, and eccentricity 0.2053 \citep{soderhjelm:1999,mason:1999,malkov:2012}. No spectroscopic companion is listed in the SB9 catalog. Thus, the companion we detect is not the stellar secondary, and presumably orbits the primary given that the flux in the spectra is heavily weighted to the brighter primary component. The third star in the system is NLTT 45161 at a distance of 9.4\arcsec\ and is 2.26 magnitudes fainter in the V band than the combined AB pair \citep{mason:2001,gould:2004}. Among FGK dwarf systems, 12\% are triple star systems \citep{raghavan:2010,tokovinin:2014}, and as of May 2021 the NASA Exoplanet Archive reports 3260 stellar systems hosting at least one confirmed exoplanet of which 41 are triple star systems, making this detection rare among the known exoplanet population.

\rftabl{orbpars} summarizes the orbital elements for the nine systems discussed here. The values listed include the first orbits from the discovery references, and our orbit. We have not added any points from other efforts to ours from CHIRON to enable direct comparisons between results. For the five stars used to check the veracity of our observing and reduction efforts with CHIRON, we find that our orbits are in good agreement with those found in the discovery papers. We note that we have been able to reach similar orbital solutions with less data compared to the discovery papers, primarily because of CHIRON's RV precision for this type of star.

\subsection{K Dwarfs for Future Low-Mass Planet Surveys}

In \rftabl{sample}, we also provide details for the 186 K dwarfs within 50 pc for which no companion down to our sensitivity and time coverage has been detected. In addition to names and epoch/equinox 2000.0 coordinates, each star's parallax from {\it Gaia} DR2, $V$ and $K$ photometry, $V-K$ color, $M_V$ magnitude, are given, as well as the number of observations and timespan of our CHIRON observations.

Note that companions further than a few AU from their primaries are beyond the sensitivity limit of our RV survey to date --- we hope to reveal most stellar companions at these separations through our high-resolution speckle survey of the same stars. The 186 stars listed should become high-priority targets for terrestrial planet searches because we now know that they do not have stellar or brown dwarf companions within a few AU. In fact, crosschecks of the NASA Exoplanet Archive as of May 2021 have revealed that none of the 186 stars have confirmed planets. However, additional checks of {\it K2} and \tess\ reveal that four --- HIP 5286, HIP 11707, HIP 12493, and HIP 74981 --- have recently been added to the \tess\ TOI list. Still, none of these stars have significant numbers of observations in the HARPS and HIRES data archives, so each remains a promising new target for deeper and more precise searches for terrestrial planets.

\section{Conclusions}

We have presented the first results of our ongoing RV survey of nearby K dwarfs with CHIRON spectrograph. Three K dwarf RV standard stars and a set of 186 stars with no detected companions have been used to determine the stability level of CHIRON over 2.5 years to be 5--20 \ms~for K dwarf stars with magnitudes of $V$ = 7--12. Previously known planets around five K dwarfs have been independently detected with CHIRON and produced orbital solutions consistent with previous efforts. We have independently confirmed a giant planet around a K dwarf initially discovered by \tess, taking data with CHIRON within a few days of the first \tess~data release. Three K dwarfs in our survey show RV variations consistent with planets of minimum masses from 0.5--0.9 \mjup~in orbital periods of 2--160 days. We provide details for 186 K dwarfs within 50 pc that do not show significant variations in RV indicative of close stellar or sub-stellar companions in orbits with periods less than a year. Vetting stars for close brown dwarf and jovian companions is a time-consuming and expensive effort in the search for terrestrial exoplanets; thus, we provide this list of K dwarfs as ideal targets for extreme precision radial velocity programs.

All of the K dwarfs in our survey are also being examined for stellar companions, as promising new multiple stellar systems show up in our data with larger RV variations, long term RV linear trends, and fully resolved orbits, we are preparing them for the next publication. Moreover, beyond the few AU regime sampled by the RV effort --- both high-resolution speckle imaging and wide-field companion searches are being done to provide a comprehensive assessment of stellar companions from 0--1000 AU, which will ultimately provide a detailed understanding of where stellar companions form around K dwarfs and what their orbital architectures look like. Because of its sensitivity, the RV survey probes beyond stellar companions to brown dwarf and massive planetary companions, providing an opportunity to evaluate architectures for all three classes of companions.

Via a carefully defined sample with systematic coverage in three spatial regimes, we will be able to reveal the results of the stellar, brown dwarf, and jovian planetary formation processes, with an ultimate expansion to the regime of terrestrial planets. In the end, we will then understand the populations of companions spanning a factor of $\sim$1000 in mass for many of the nearest K dwarfs to the Sun.


\section{Acknowledgments}

This work has been supported by the National Science Foundation via grants AST-1517413 and AST-1910130. 
We are indebted to members of the SMARTS Consortium and NSF’s National Optical-Infrared Astronomy Research Laboratory, particularly the staff at CTIO, for efforts to keep the 1.5m and the CHIRON spectrograph in operation. 
At CTIO, Rodrigo Hinojosa, Roberto Aviles, Hernan Tirado, Manuel Hernandez, David Rojas, Javier Rojas, Humberto Orrego, Esteban Parkes, Marco Bonati, Peter Moore, Nicole David, Mauricio Rojas, Andrei Tokovinin, Steve Heathcote, Ximena Herreros, Luz Pinto, Rodrigo Hernandez, Jacqueline Seron, Fernando Cortes, Carlos Corco, Alberto Miranda, Jorge Briones, Carlos Correa, Nelson Ogalde, Cristian Diaz, Juan Andrade, Jorge Lopez, Pedro Ramos, Alfonso Rojas, Hector Pasten, Juan Paleo, Marco Nunez, and so many others. We also want to thank Eric Mamajek for his insightful comments on this work.
This research has made use of the NASA Exoplanet Archive, which is operated by the California Institute of Technology, under contract with the National Aeronautics and Space Administration under the Exoplanet Exploration Program.
This publication makes use of data products from the Two Micron All Sky Survey, which is a joint project of the University of Massachusetts and the Infrared Processing and Analysis Center/California Institute of Technology, funded by the National Aeronautics and Space Administration and the National Science Foundation.
This work has made use of data from the European Space Agency's $Hipparcos$ mission and {\it Gaia} (\href{https://www.cosmos.esa.int/gaia}{cosmos.esa.int/gaia}) missions, the latter processed by the {\it Gaia} Data Processing and Analysis Consortium (DPAC,
\href{https://www.cosmos.esa.int/web/gaia/dpac/consortium}{cosmos.esa.int/web/gaia/dpac/consortium}). Funding
for the DPAC has been provided by national institutions, in particular
the institutions participating in the {\it Gaia} Multilateral Agreement.

\facilities{CTIO:1.5m, TESS, Exoplanet Archive}




\appendix
\restartappendixnumbering
\section{Additional tables}

\startlongtable
\begin{deluxetable}{crr}
\setlength{\tabcolsep}{0.2in}
\tablewidth{0pt}
\tabletypesize{\small}
\tablecaption{\label{tab:rvstds} CHIRON Radial Velocities of K Dwarf Radial Velocity Standards. $(^\star)$ marks the reference epoch.}
\tablehead{\colhead{${BJD} - 2450000$} &
           \colhead{$RV$}              & 
           \colhead{$\sigma_{RV}$}    \\
           \colhead{(days)}            &
           \colhead{(\ms)}             &
           \colhead{(\ms)}}
\startdata
\multicolumn{3}{c}{}
\\
\hline
\multicolumn{3}{c}{HIP003535}
\\
\hline
7937.93779      &          3.6   &       4.5  \\ %
7952.91229      &         31.3   &       6.9  \\ %
7953.91139      &         12.0   &       3.6  \\ %
7955.89438      &         19.4   &       4.3  \\ %
7956.89693      &         50.4   &       7.1  \\ %
7957.88705      &          4.0   &       3.4  \\
\enddata
\end{deluxetable}

\startlongtable
\begin{deluxetable}{crr}
\setlength{\tabcolsep}{0.2in}
\tablewidth{0pt}
\tabletypesize{\small}
\tablecaption{\label{tab:rvs} CHIRON Radial Velocities of Known and New Planet Candidates. $(^\star)$ marks the reference epoch.}
\tablehead{\colhead{$BJD - 2450000$}   &
           \colhead{$RV$}              & 
           \colhead{$\sigma_{RV}$}     \\
           \colhead{(days)}            &
           \colhead{(\ms)}             &
           \colhead{(\ms)}}
\startdata
\multicolumn{3}{c}{}
\\
\hline
\multicolumn{3}{c}{HIP002350}
\\
\hline
7951.89442      &         77.2   &      14.5  \\ %
7952.90012      &        -36.1   &      10.6  \\ %
7953.89963      &        -33.5   &       9.0  \\ %
7955.88144      &         -3.8   &      10.0  \\ %
7956.88370      &        -41.0   &       9.4  \\ %
7957.87499      &         38.4   &      10.2  \\
\enddata
\end{deluxetable}

\clearpage
\startlongtable
\begin{deluxetable*}{rrrrrrrrrrr}
\setlength{\tabcolsep}{0.1in}
\tablewidth{0pt}
\tabletypesize{\small}
\tablecaption{\label{tab:sample} Sample of 198 K dwarfs with RV curves}
\tablehead{\colhead{HIP}             &
           \colhead{RA}              & 
           \colhead{DEC}             &
           \colhead{Parallax}        &
           \colhead{$V$}             & 
           \colhead{$K$}             &
           \colhead{$V-K$}           &
           \colhead{$M_{V}$}         &
           \colhead{MAD$_{RV}$}      &
           \colhead{\# obs.}         &
           \colhead{Time span}      \\
           \colhead{}                &
           \colhead{hh:mm:ss.sss}    &
           \colhead{$^{\circ}$:':''} &
           \colhead{mas}             &
           \colhead{mag}             &
           \colhead{mag}             &
           \colhead{mag}             &
           \colhead{mag}             &
           \colhead{\ms}             &
           \colhead{}                &
           \colhead{days}            
           }
\startdata
\multicolumn{11}{c}{}
\\
\hline
\multicolumn{11}{c}{K dwarf RV standards}
\\
\hline
  3535 &  00:45:04.894  &  +01:47:07.88  &  46.37  &   8.01  &  5.74  &  2.26  &  6.34  &   15.2  &   77  &  892  \\ %
 58345 &  11:57:56.207  &  -27:42:25.37  &  98.45  &   6.97  &  4.53  &  2.44  &  6.93  &    9.8  &   62  &  733  \\ %
 73184 &  14:57:28.001  &  -21:24:55.71  & 171.22  &   5.75  &  3.80  &  1.94  &  6.91  &   10.3  &   77  &   56  \\ %
\multicolumn{11}{c}{}
\\
\hline
\multicolumn{11}{c}{K dwarfs with known planets}
\\
\hline
  2350 &  00:29:59.873  &  -05:45:50.40  &  20.03  &   9.37  &  7.31  &  2.06  &  5.88  &   39.5  &   24  &   34  \\
\enddata
\end{deluxetable*}

\bibliography{main.bbl}

\begin{thebibliography}{}
\expandafter\ifx\csname natexlab\endcsname\relax\def\natexlab#1{#1}\fi
\providecommand{\url}[1]{\href{#1}{#1}}
\providecommand{\dodoi}[1]{doi:~\href{http://doi.org/#1}{\nolinkurl{#1}}}
\providecommand{\doeprint}[1]{\href{http://ascl.net/#1}{\nolinkurl{http://ascl.net/#1}}}
\providecommand{\doarXiv}[1]{\href{https://arxiv.org/abs/#1}{\nolinkurl{https://arxiv.org/abs/#1}}}

\bibitem[{{Baglin} {et~al.}(2006){Baglin}, {Auvergne}, {Barge}, {Deleuil},
  {Catala}, {Michel}, {Weiss}, \& {COROT Team}}]{baglin:2006}
{Baglin}, A., {Auvergne}, M., {Barge}, P., {et~al.} 2006, in ESA Special
  Publication, Vol. 1306, The CoRoT Mission Pre-Launch Status - Stellar
  Seismology and Planet Finding, ed. M.~{Fridlund}, A.~{Baglin}, J.~{Lochard},
  \& L.~{Conroy}, 33

\bibitem[{{Bakos} {et~al.}(2006){Bakos}, {P{\'a}l}, {Latham}, {Noyes}, \&
  {Stefanik}}]{bakos:2006}
{Bakos}, G.~{\'A}., {P{\'a}l}, A., {Latham}, D.~W., {Noyes}, R.~W., \&
  {Stefanik}, R.~P. 2006, \apjl, 641, L57, \dodoi{10.1086/503671}

\bibitem[{{Baranec} {et~al.}(2012){Baranec}, {Riddle}, {Ramaprakash}, {Law},
  {Tendulkar}, {Kulkarni}, {Dekany}, {Bui}, {Davis}, {Burse}, {Das},
  {Hildebrandt}, {Punnadi}, \& {Smith}}]{baranec:2012}
{Baranec}, C., {Riddle}, R., {Ramaprakash}, A.~N., {et~al.} 2012, in Society of
  Photo-Optical Instrumentation Engineers (SPIE) Conference Series, Vol. 8447,
  Adaptive Optics Systems III, ed. B.~L. {Ellerbroek}, E.~{Marchetti}, \& J.-P.
  {V{\'e}ran}, 844704, \dodoi{10.1117/12.924867}

\bibitem[{{Baranne} {et~al.}(1996){Baranne}, {Queloz}, {Mayor}, {Adrianzyk},
  {Knispel}, {Kohler}, {Lacroix}, {Meunier}, {Rimbaud}, \&
  {Vin}}]{baranne:1996}
{Baranne}, A., {Queloz}, D., {Mayor}, M., {et~al.} 1996, \aaps, 119, 373

\bibitem[{{Bernstein} {et~al.}(2003){Bernstein}, {Shectman}, {Gunnels},
  {Mochnacki}, \& {Athey}}]{bernstein:2003}
{Bernstein}, R., {Shectman}, S.~A., {Gunnels}, S.~M., {Mochnacki}, S., \&
  {Athey}, A.~E. 2003, in Society of Photo-Optical Instrumentation Engineers
  (SPIE) Conference Series, Vol. 4841, Instrument Design and Performance for
  Optical/Infrared Ground-based Telescopes, ed. M.~{Iye} \& A.~F.~M.
  {Moorwood}, 1694--1704, \dodoi{10.1117/12.461502}

\bibitem[{{Borucki} {et~al.}(2010){Borucki}, {Koch}, {Basri}, {Batalha},
  {Brown}, {Caldwell}, {Caldwell}, {Christensen-Dalsgaard}, {Cochran},
  {DeVore}, {Dunham}, {Dupree}, {Gautier}, {Geary}, {Gilliland}, {Gould},
  {Howell}, {Jenkins}, {Kondo}, {Latham}, {Marcy}, {Meibom}, {Kjeldsen},
  {Lissauer}, {Monet}, {Morrison}, {Sasselov}, {Tarter}, {Boss}, {Brownlee},
  {Owen}, {Buzasi}, {Charbonneau}, {Doyle}, {Fortney}, {Ford}, {Holman},
  {Seager}, {Steffen}, {Welsh}, {Rowe}, {Anderson}, {Buchhave}, {Ciardi},
  {Walkowicz}, {Sherry}, {Horch}, {Isaacson}, {Everett}, {Fischer}, {Torres},
  {Johnson}, {Endl}, {MacQueen}, {Bryson}, {Dotson}, {Haas}, {Kolodziejczak},
  {Van Cleve}, {Chandrasekaran}, {Twicken}, {Quintana}, {Clarke}, {Allen},
  {Li}, {Wu}, {Tenenbaum}, {Verner}, {Bruhweiler}, {Barnes}, \&
  {Prsa}}]{borucki:2010}
{Borucki}, W.~J., {Koch}, D., {Basri}, G., {et~al.} 2010, Science, 327, 977,
  \dodoi{10.1126/science.1185402}

\bibitem[{{Bouchy} {et~al.}(2005){Bouchy}, {Udry}, {Mayor}, {Moutou}, {Pont},
  {Iribarne}, {da Silva}, {Ilovaisky}, {Queloz}, {Santos}, {S{\'e}gransan}, \&
  {Zucker}}]{bouchy:2005}
{Bouchy}, F., {Udry}, S., {Mayor}, M., {et~al.} 2005, \aap, 444, L15,
  \dodoi{10.1051/0004-6361:200500201}

\bibitem[{{Brewer} {et~al.}(2014){Brewer}, {Giguere}, \&
  {Fischer}}]{brewer:2014}
{Brewer}, J.~M., {Giguere}, M., \& {Fischer}, D.~A. 2014, \pasp, 126, 48,
  \dodoi{10.1086/674723}

\bibitem[{{Butler} {et~al.}(2017){Butler}, {Vogt}, {Laughlin}, {Burt},
  {Rivera}, {Tuomi}, {Teske}, {Arriagada}, {Diaz}, {Holden}, \&
  {Keiser}}]{butler:2017}
{Butler}, R.~P., {Vogt}, S.~S., {Laughlin}, G., {et~al.} 2017, \aj, 153, 208,
  \dodoi{10.3847/1538-3881/aa66ca}

\bibitem[{{Crane} {et~al.}(2006){Crane}, {Shectman}, \& {Butler}}]{crane:2006}
{Crane}, J.~D., {Shectman}, S.~A., \& {Butler}, R.~P. 2006, in Society of
  Photo-Optical Instrumentation Engineers (SPIE) Conference Series, Vol. 6269,
  Society of Photo-Optical Instrumentation Engineers (SPIE) Conference Series,
  626931, \dodoi{10.1117/12.672339}

\bibitem[{{Cuntz} \& {Guinan}(2016)}]{cuntz:2016}
{Cuntz}, M., \& {Guinan}, E.~F. 2016, \apj, 827, 79,
  \dodoi{10.3847/0004-637X/827/1/79}

\bibitem[{{Evans} {et~al.}(2018){Evans}, {Riello}, {De Angeli}, {Carrasco},
  {Montegriffo}, {Fabricius}, {Jordi}, {Palaversa}, {Diener}, {Busso},
  {Cacciari}, {van Leeuwen}, {Burgess}, {Davidson}, {Harrison}, {Hodgkin},
  {Pancino}, {Richards}, {Altavilla}, {Balaguer-N{\'u}{\~n}ez}, {Barstow},
  {Bellazzini}, {Brown}, {Castellani}, {Cocozza}, {De Luise}, {Delgado},
  {Ducourant}, {Galleti}, {Gilmore}, {Giuffrida}, {Holl}, {Kewley}, {Koposov},
  {Marinoni}, {Marrese}, {Osborne}, {Piersimoni}, {Portell}, {Pulone},
  {Ragaini}, {Sanna}, {Terrett}, {Walton}, {Wevers}, \&
  {Wyrzykowski}}]{evans:2018}
{Evans}, D.~W., {Riello}, M., {De Angeli}, F., {et~al.} 2018, \aap, 616, A4,
  \dodoi{10.1051/0004-6361/201832756}

\bibitem[{{Ge} {et~al.}(2006){Ge}, {van Eyken}, {Mahadevan}, {DeWitt}, {Kane},
  {Cohen}, {Vand en Heuvel}, {Fleming}, {Guo}, {Henry}, {Schneider}, {Ramsey},
  {Wittenmyer}, {Endl}, {Cochran}, {Ford}, {Mart{\'\i}n}, {Israelian},
  {Valenti}, \& {Montes}}]{ge:2006}
{Ge}, J., {van Eyken}, J., {Mahadevan}, S., {et~al.} 2006, \apj, 648, 683,
  \dodoi{10.1086/505699}

\bibitem[{{Giguere} {et~al.}(2015){Giguere}, {Fischer}, {Payne}, {Brewer},
  {Johnson}, {Howard}, \& {Isaacson}}]{giguere:2015}
{Giguere}, M.~J., {Fischer}, D.~A., {Payne}, M.~J., {et~al.} 2015, \apj, 799,
  89, \dodoi{10.1088/0004-637X/799/1/89}

\bibitem[{{Giguere} {et~al.}(2016){Giguere}, {Fischer}, {Zhang}, {Matthews},
  {Cameron}, \& {Henry}}]{giguere:2016}
{Giguere}, M.~J., {Fischer}, D.~A., {Zhang}, C. X.~Y., {et~al.} 2016, \apj,
  824, 150, \dodoi{10.3847/0004-637X/824/2/150}

\bibitem[{{Gilbert} {et~al.}(2018){Gilbert}, {Bergmann}, {Bloxham}, {Boz},
  {Brookfield}, {Carkic}, {Carter}, {Case}, {Churilov}, {Ellis}, {Gausachs},
  {Gers}, {Gray}, {Herrald}, {Ireland}, {Jones}, {Kripak}, {Lawrence},
  {O'Brien}, {Price}, {Robertson}, {Schwab}, {Tinney}, {Vaccarella}, {Vest},
  {Wright}, \& {Zhelem}}]{gilbert:2018}
{Gilbert}, J., {Bergmann}, C., {Bloxham}, G., {et~al.} 2018, in Society of
  Photo-Optical Instrumentation Engineers (SPIE) Conference Series, Vol. 10702,
  Ground-based and Airborne Instrumentation for Astronomy VII, 107020Y,
  \dodoi{10.1117/12.2312399}

\bibitem[{{Gould} \& {Chanam{\'e}}(2004)}]{gould:2004}
{Gould}, A., \& {Chanam{\'e}}, J. 2004, \apjs, 150, 455, \dodoi{10.1086/381147}

\bibitem[{{Gray} \& {Corbally}(2009)}]{gray:2009}
{Gray}, R.~O., \& {Corbally}, Christopher, J. 2009, {Stellar Spectral
  Classification}

\bibitem[{{Guilluy} {et~al.}(2019){Guilluy}, {Sozzetti}, {Brogi}, {Bonomo},
  {Giacobbe}, {Claudi}, \& {Benatti}}]{guilluy:2019}
{Guilluy}, G., {Sozzetti}, A., {Brogi}, M., {et~al.} 2019, \aap, 625, A107,
  \dodoi{10.1051/0004-6361/201834615}

\bibitem[{{Guilluy} {et~al.}(2020){Guilluy}, {Andretta}, {Borsa}, {Giacobbe},
  {Sozzetti}, {Covino}, {Bourrier}, {Fossati}, {Bonomo}, {Esposito},
  {Giampapa}, {Harutyunyan}, {Rainer}, {Brogi}, {Bruno}, {Claudi}, {Frustagli},
  {Lanza}, {Mancini}, {Pino}, {Poretti}, {Scandariato}, {Affer}, {Baffa},
  {Baruffolo}, {Benatti}, {Biazzo}, {Bignamini}, {Boschin}, {Carleo},
  {Cecconi}, {Cosentino}, {Damasso}, {Desidera}, {Falcini}, {Martinez
  Fiorenzano}, {Ghedina}, {Gonz{\'a}lez-{\'A}lvarez}, {Guerra}, {Hernandez},
  {Leto}, {Maggio}, {Malavolta}, {Maldonado}, {Micela}, {Molinari},
  {Nascimbeni}, {Pagano}, {Pedani}, {Piotto}, \& {Reiners}}]{guilluy:2020}
{Guilluy}, G., {Andretta}, V., {Borsa}, F., {et~al.} 2020, \aap, 639, A49,
  \dodoi{10.1051/0004-6361/202037644}

\bibitem[{{Halbwachs} {et~al.}(2005){Halbwachs}, {Mayor}, \&
  {Udry}}]{halbwachs:2005}
{Halbwachs}, J.~L., {Mayor}, M., \& {Udry}, S. 2005, \aap, 431, 1129,
  \dodoi{10.1051/0004-6361:20041219}

\bibitem[{{Henry} {et~al.}(2002){Henry}, {Donahue}, \& {Baliunas}}]{henry:2002}
{Henry}, G.~W., {Donahue}, R.~A., \& {Baliunas}, S.~L. 2002, \apjl, 577, L111,
  \dodoi{10.1086/344291}

\bibitem[{{Henry} {et~al.}(1997){Henry}, {Ianna}, {Kirkpatrick}, \&
  {Jahreiss}}]{henry:1997}
{Henry}, T.~J., {Ianna}, P.~A., {Kirkpatrick}, J.~D., \& {Jahreiss}, H. 1997,
  \aj, 114, 388, \dodoi{10.1086/118482}

\bibitem[{{Henry} {et~al.}(2006){Henry}, {Jao}, {Subasavage}, {Beaulieu},
  {Ianna}, {Costa}, \& {M{\'e}ndez}}]{henry:2006}
{Henry}, T.~J., {Jao}, W.-C., {Subasavage}, J.~P., {et~al.} 2006, \aj, 132,
  2360, \dodoi{10.1086/508233}

\bibitem[{{Henry} \& {McCarthy}(1993)}]{henry:1993}
{Henry}, T.~J., \& {McCarthy}, Donald~W., J. 1993, \aj, 106, 773,
  \dodoi{10.1086/116685}

\bibitem[{{Henry} {et~al.}(2018){Henry}, {Jao}, {Winters}, {Dieterich},
  {Finch}, {Ianna}, {Riedel}, {Silverstein}, {Subasavage}, \&
  {Vrijmoet}}]{henry:2018}
{Henry}, T.~J., {Jao}, W.-C., {Winters}, J.~G., {et~al.} 2018, \aj, 155, 265,
  \dodoi{10.3847/1538-3881/aac262}

\bibitem[{{Hinkel} {et~al.}(2015){Hinkel}, {Kane}, {Henry}, {Feng}, {Boyajian},
  {Wright}, {Fischer}, \& {Howard}}]{hinkel:2015}
{Hinkel}, N.~R., {Kane}, S.~R., {Henry}, G.~W., {et~al.} 2015, \apj, 803, 8,
  \dodoi{10.1088/0004-637X/803/1/8}

\bibitem[{{H{\o}g} {et~al.}(2000){H{\o}g}, {Fabricius}, {Makarov}, {Urban},
  {Corbin}, {Wycoff}, {Bastian}, {Schwekendiek}, \& {Wicenec}}]{hog:2000}
{H{\o}g}, E., {Fabricius}, C., {Makarov}, V.~V., {et~al.} 2000, \aap, 355, L27

\bibitem[{{Horch} {et~al.}(2009){Horch}, {Veillette}, {Baena Gall{\'e}},
  {Shah}, {O'Rielly}, \& {van Altena}}]{horch:2009}
{Horch}, E.~P., {Veillette}, D.~R., {Baena Gall{\'e}}, R., {et~al.} 2009, \aj,
  137, 5057, \dodoi{10.1088/0004-6256/137/6/5057}

\bibitem[{{Howell} {et~al.}(2014){Howell}, {Sobeck}, {Haas}, {Still},
  {Barclay}, {Mullally}, {Troeltzsch}, {Aigrain}, {Bryson}, {Caldwell},
  {Chaplin}, {Cochran}, {Huber}, {Marcy}, {Miglio}, {Najita}, {Smith},
  {Twicken}, \& {Fortney}}]{howell:2014}
{Howell}, S.~B., {Sobeck}, C., {Haas}, M., {et~al.} 2014, \pasp, 126, 398,
  \dodoi{10.1086/676406}

\bibitem[{{Kaufer} \& {Pasquini}(1998)}]{kaufer:1998}
{Kaufer}, A., \& {Pasquini}, L. 1998, in Society of Photo-Optical
  Instrumentation Engineers (SPIE) Conference Series, Vol. 3355, Optical
  Astronomical Instrumentation, ed. S.~{D'Odorico}, 844--854,
  \dodoi{10.1117/12.316798}

\bibitem[{{Latham} {et~al.}(1989){Latham}, {Mazeh}, {Stefanik}, {Mayor}, \&
  {Burki}}]{latham:1989}
{Latham}, D.~W., {Mazeh}, T., {Stefanik}, R.~P., {Mayor}, M., \& {Burki}, G.
  1989, \nat, 339, 38, \dodoi{10.1038/339038a0}

\bibitem[{{Malkov} {et~al.}(2012){Malkov}, {Tamazian}, {Docobo}, \&
  {Chulkov}}]{malkov:2012}
{Malkov}, O.~Y., {Tamazian}, V.~S., {Docobo}, J.~A., \& {Chulkov}, D.~A. 2012,
  \aap, 546, A69, \dodoi{10.1051/0004-6361/201219774}

\bibitem[{{Mamajek}(2012)}]{mamajek:2012}
{Mamajek}, E.~E. 2012, arXiv e-prints, arXiv:1210.1616.
\newblock \doarXiv{1210.1616}

\bibitem[{{Mason} {et~al.}(1999){Mason}, {Douglass}, \&
  {Hartkopf}}]{mason:1999}
{Mason}, B.~D., {Douglass}, G.~G., \& {Hartkopf}, W.~I. 1999, \aj, 117, 1023,
  \dodoi{10.1086/300748}

\bibitem[{{Mason} {et~al.}(2001){Mason}, {Wycoff}, {Hartkopf}, {Douglass}, \&
  {Worley}}]{mason:2001}
{Mason}, B.~D., {Wycoff}, G.~L., {Hartkopf}, W.~I., {Douglass}, G.~G., \&
  {Worley}, C.~E. 2001, \aj, 122, 3466, \dodoi{10.1086/323920}

\bibitem[{{Mayor} \& {Queloz}(1995)}]{mayor:1995}
{Mayor}, M., \& {Queloz}, D. 1995, \nat, 378, 355, \dodoi{10.1038/378355a0}

\bibitem[{{Mayor} {et~al.}(2003){Mayor}, {Pepe}, {Queloz}, {Bouchy},
  {Rupprecht}, {Lo Curto}, {Avila}, {Benz}, {Bertaux}, {Bonfils}, {Dall},
  {Dekker}, {Delabre}, {Eckert}, {Fleury}, {Gilliotte}, {Gojak}, {Guzman},
  {Kohler}, {Lizon}, {Longinotti}, {Lovis}, {Megevand}, {Pasquini}, {Reyes},
  {Sivan}, {Sosnowska}, {Soto}, {Udry}, {van Kesteren}, {Weber}, \&
  {Weilenmann}}]{mayor:2003}
{Mayor}, M., {Pepe}, F., {Queloz}, D., {et~al.} 2003, The Messenger, 114, 20

\bibitem[{{Mayor} {et~al.}(2011){Mayor}, {Marmier}, {Lovis}, {Udry},
  {S{\'e}gransan}, {Pepe}, {Benz}, {Bertaux}, {Bouchy}, {Dumusque}, {Lo Curto},
  {Mordasini}, {Queloz}, \& {Santos}}]{mayor:2011}
{Mayor}, M., {Marmier}, M., {Lovis}, C., {et~al.} 2011, arXiv e-prints,
  arXiv:1109.2497.
\newblock \doarXiv{1109.2497}

\bibitem[{{Meschiari} {et~al.}(2009){Meschiari}, {Wolf}, {Rivera}, {Laughlin},
  {Vogt}, \& {Butler}}]{meschiari:2009}
{Meschiari}, S., {Wolf}, A.~S., {Rivera}, E., {et~al.} 2009, \pasp, 121, 1016,
  \dodoi{10.1086/605730}

\bibitem[{{Moutou} {et~al.}(2005){Moutou}, {Mayor}, {Bouchy}, {Lovis}, {Pepe},
  {Queloz}, {Santos}, {Udry}, {Benz}, {Lo Curto}, {Naef}, {S{\'e}gransan}, \&
  {Sivan}}]{moutou:2005}
{Moutou}, C., {Mayor}, M., {Bouchy}, F., {et~al.} 2005, \aap, 439, 367,
  \dodoi{10.1051/0004-6361:20052826}

\bibitem[{{Munari} \& {Walter}(2016)}]{munari:2016}
{Munari}, U., \& {Walter}, F.~M. 2016, \mnras, 455, L57,
  \dodoi{10.1093/mnrasl/slv146}

\bibitem[{{Nielsen} {et~al.}(2020){Nielsen}, {Brahm}, {Bouchy}, {Espinoza},
  {Turner}, {Rappaport}, {Pearce}, {Ricker}, {Vanderspek}, {Latham}, {Seager},
  {Winn}, {Jenkins}, {Acton}, {Bakos}, {Barclay}, {Barkaoui}, {Bhatti},
  {Brice{\~n}o}, {Bryant}, {Burleigh}, {Ciardi}, {Collins}, {Collins}, {Cooke},
  {Csubry}, {dos Santos}, {Eigm{\"u}ller}, {Fausnaugh}, {Gan}, {Gillon},
  {Goad}, {Guerrero}, {Hagelberg}, {Hart}, {Henning}, {Huang}, {Jehin},
  {Jenkins}, {Jord{\'a}n}, {Kielkopf}, {Kossakowski}, {Lavie}, {Law}, {Lendl},
  {de Leon}, {Lovis}, {Mann}, {Marmier}, {McCormac}, {Mori}, {Moyano},
  {Narita}, {Osip}, {Otegi}, {Pepe}, {Pozuelos}, {Raynard}, {Relles}, {Sarkis},
  {S{\'e}gransan}, {Seidel}, {Shporer}, {Stalport}, {Stockdale}, {Suc},
  {Tamura}, {Tan}, {Tilbrook}, {Ting}, {Trifonov}, {Udry}, {Vanderburg},
  {Wheatley}, {Wingham}, {Zhan}, \& {Ziegler}}]{nielsen:2020}
{Nielsen}, L.~D., {Brahm}, R., {Bouchy}, F., {et~al.} 2020, \aap, 639, A76,
  \dodoi{10.1051/0004-6361/202037941}

\bibitem[{{Oh} {et~al.}(2017){Oh}, {Price-Whelan}, {Hogg}, {Morton}, \&
  {Spergel}}]{oh:2017}
{Oh}, S., {Price-Whelan}, A.~M., {Hogg}, D.~W., {Morton}, T.~D., \& {Spergel},
  D.~N. 2017, \aj, 153, 257, \dodoi{10.3847/1538-3881/aa6ffd}

\bibitem[{{Oliva} {et~al.}(2006){Oliva}, {Origlia}, {Baffa}, {Biliotti},
  {Bruno}, {D'Amato}, {Del Vecchio}, {Falcini}, {Gennari}, {Ghinassi}, {Giani},
  {Gonzalez}, {Leone}, {Lolli}, {Lodi}, {Maiolino}, {Mannucci}, {Marcucci},
  {Mochi}, {Montegriffo}, {Rossetti}, {Scuderi}, \& {Sozzi}}]{oliva:2006}
{Oliva}, E., {Origlia}, L., {Baffa}, C., {et~al.} 2006, in Society of
  Photo-Optical Instrumentation Engineers (SPIE) Conference Series, Vol. 6269,
  Society of Photo-Optical Instrumentation Engineers (SPIE) Conference Series,
  ed. I.~S. {McLean} \& M.~{Iye}, 626919, \dodoi{10.1117/12.670006}

\bibitem[{{Piskunov} \& {Valenti}(2002)}]{piskunov:2002}
{Piskunov}, N.~E., \& {Valenti}, J.~A. 2002, \aap, 385, 1095,
  \dodoi{10.1051/0004-6361:20020175}

\bibitem[{{Pourbaix} {et~al.}(2004){Pourbaix}, {Tokovinin}, {Batten}, {Fekel},
  {Hartkopf}, {Levato}, {Morrell}, {Torres}, \& {Udry}}]{pourbaix:2004}
{Pourbaix}, D., {Tokovinin}, A.~A., {Batten}, A.~H., {et~al.} 2004, \aap, 424,
  727, \dodoi{10.1051/0004-6361:20041213}

\bibitem[{{Raghavan} {et~al.}(2010){Raghavan}, {McAlister}, {Henry}, {Latham},
  {Marcy}, {Mason}, {Gies}, {White}, \& {ten Brummelaar}}]{raghavan:2010}
{Raghavan}, D., {McAlister}, H.~A., {Henry}, T.~J., {et~al.} 2010, \apjs, 190,
  1, \dodoi{10.1088/0067-0049/190/1/1}

\bibitem[{{Redfield} {et~al.}(2008){Redfield}, {Endl}, {Cochran}, \&
  {Koesterke}}]{redfield:2008}
{Redfield}, S., {Endl}, M., {Cochran}, W.~D., \& {Koesterke}, L. 2008, \apjl,
  673, L87, \dodoi{10.1086/527475}

\bibitem[{{Ricker} {et~al.}(2014){Ricker}, {Winn}, {Vanderspek}, {Latham},
  {Bakos}, {Bean}, {Berta-Thompson}, {Brown}, {Buchhave}, {Butler}, {Butler},
  {Chaplin}, {Charbonneau}, {Christensen-Dalsgaard}, {Clampin}, {Deming},
  {Doty}, {De Lee}, {Dressing}, {Dunham}, {Endl}, {Fressin}, {Ge}, {Henning},
  {Holman}, {Howard}, {Ida}, {Jenkins}, {Jernigan}, {Johnson}, {Kaltenegger},
  {Kawai}, {Kjeldsen}, {Laughlin}, {Levine}, {Lin}, {Lissauer}, {MacQueen},
  {Marcy}, {McCullough}, {Morton}, {Narita}, {Paegert}, {Palle}, {Pepe},
  {Pepper}, {Quirrenbach}, {Rinehart}, {Sasselov}, {Sato}, {Seager},
  {Sozzetti}, {Stassun}, {Sullivan}, {Szentgyorgyi}, {Torres}, {Udry}, \&
  {Villasenor}}]{ricker:2014}
{Ricker}, G.~R., {Winn}, J.~N., {Vanderspek}, R., {et~al.} 2014, in Society of
  Photo-Optical Instrumentation Engineers (SPIE) Conference Series, Vol. 9143,
  Space Telescopes and Instrumentation 2014: Optical, Infrared, and Millimeter
  Wave, 914320, \dodoi{10.1117/12.2063489}

\bibitem[{{Riddle} {et~al.}(2015){Riddle}, {Tokovinin}, {Mason}, {Hartkopf},
  {Roberts}, {Baranec}, {Law}, {Bui}, {Burse}, {Das}, {Dekany}, {Kulkarni},
  {Punnadi}, {Ramaprakash}, \& {Tendulkar}}]{riddle:2015}
{Riddle}, R.~L., {Tokovinin}, A., {Mason}, B.~D., {et~al.} 2015, \apj, 799, 4,
  \dodoi{10.1088/0004-637X/799/1/4}

\bibitem[{{Roberts} {et~al.}(2015){Roberts}, {Tokovinin}, {Mason}, {Riddle},
  {Hartkopf}, {Law}, \& {Baranec}}]{roberts:2015}
{Roberts}, Lewis~C., J., {Tokovinin}, A., {Mason}, B.~D., {et~al.} 2015, \aj,
  149, 118, \dodoi{10.1088/0004-6256/149/4/118}

\bibitem[{{Santos} {et~al.}(2000){Santos}, {Mayor}, {Naef}, {Pepe}, {Queloz},
  {Udry}, {Burnet}, \& {Revaz}}]{santos:2000}
{Santos}, N.~C., {Mayor}, M., {Naef}, D., {et~al.} 2000, \aap, 356, 599

\bibitem[{{Santos} {et~al.}(2003){Santos}, {Udry}, {Mayor}, {Naef}, {Pepe},
  {Queloz}, {Burki}, {Cramer}, \& {Nicolet}}]{santos:2003}
{Santos}, N.~C., {Udry}, S., {Mayor}, M., {et~al.} 2003, \aap, 406, 373,
  \dodoi{10.1051/0004-6361:20030776}

\bibitem[{{Shaya} \& {Olling}(2011)}]{shaya:2011}
{Shaya}, E.~J., \& {Olling}, R.~P. 2011, \apjs, 192, 2,
  \dodoi{10.1088/0067-0049/192/1/2}

\bibitem[{{Skrutskie} {et~al.}(2006){Skrutskie}, {Cutri}, {Stiening},
  {Weinberg}, {Schneider}, {Carpenter}, {Beichman}, {Capps}, {Chester},
  {Elias}, {Huchra}, {Liebert}, {Lonsdale}, {Monet}, {Price}, {Seitzer},
  {Jarrett}, {Kirkpatrick}, {Gizis}, {Howard}, {Evans}, {Fowler}, {Fullmer},
  {Hurt}, {Light}, {Kopan}, {Marsh}, {McCallon}, {Tam}, {Van Dyk}, \&
  {Wheelock}}]{skrutskie:2006}
{Skrutskie}, M.~F., {Cutri}, R.~M., {Stiening}, R., {et~al.} 2006, \aj, 131,
  1163, \dodoi{10.1086/498708}

\bibitem[{{S{\"o}derhjelm}(1999)}]{soderhjelm:1999}
{S{\"o}derhjelm}, S. 1999, \aap, 341, 121

\bibitem[{{Subasavage} {et~al.}(2010){Subasavage}, {Bailyn}, {Smith}, {Henry},
  {Walter}, \& {Buxton}}]{subasavage:2010}
{Subasavage}, J.~P., {Bailyn}, C.~D., {Smith}, R.~C., {et~al.} 2010, in Society
  of Photo-Optical Instrumentation Engineers (SPIE) Conference Series, Vol.
  7737, Observatory Operations: Strategies, Processes, and Systems III, 77371C,
  \dodoi{10.1117/12.859145}

\bibitem[{{Tala} {et~al.}(2014){Tala}, {Berdja}, {Jones}, {Vanzi}, {Ropert},
  {Flores}, \& {Viscasillas}}]{tala:2014}
{Tala}, M., {Berdja}, A., {Jones}, M., {et~al.} 2014, in Society of
  Photo-Optical Instrumentation Engineers (SPIE) Conference Series, Vol. 9147,
  Ground-based and Airborne Instrumentation for Astronomy V, 914789,
  \dodoi{10.1117/12.2056551}

\bibitem[{{Tokovinin}(2014)}]{tokovinin:2014}
{Tokovinin}, A. 2014, \aj, 147, 87, \dodoi{10.1088/0004-6256/147/4/87}

\bibitem[{{Tokovinin} {et~al.}(2013){Tokovinin}, {Fischer}, {Bonati},
  {Giguere}, {Moore}, {Schwab}, {Spronck}, \& {Szymkowiak}}]{tokovinin:2013}
{Tokovinin}, A., {Fischer}, D.~A., {Bonati}, M., {et~al.} 2013, \pasp, 125,
  1336, \dodoi{10.1086/674012}

\bibitem[{{Udry} {et~al.}(2000){Udry}, {Mayor}, {Naef}, {Pepe}, {Queloz},
  {Santos}, {Burnet}, {Confino}, \& {Melo}}]{udry:2000}
{Udry}, S., {Mayor}, M., {Naef}, D., {et~al.} 2000, \aap, 356, 590

\bibitem[{{van Leeuwen}(2007)}]{vanleeuwen:2007}
{van Leeuwen}, F. 2007, \aap, 474, 653, \dodoi{10.1051/0004-6361:20078357}

\bibitem[{{Vogt} {et~al.}(1994){Vogt}, {Allen}, {Bigelow}, {Bresee}, {Brown},
  {Cantrall}, {Conrad}, {Couture}, {Delaney}, {Epps}, {Hilyard}, {Hilyard},
  {Horn}, {Jern}, {Kanto}, {Keane}, {Kibrick}, {Lewis}, {Osborne},
  {Pardeilhan}, {Pfister}, {Ricketts}, {Robinson}, {Stover}, {Tucker}, {Ward},
  \& {Wei}}]{vogt:1994}
{Vogt}, S.~S., {Allen}, S.~L., {Bigelow}, B.~C., {et~al.} 1994, in Society of
  Photo-Optical Instrumentation Engineers (SPIE) Conference Series, Vol. 2198,
  Instrumentation in Astronomy VIII, ed. D.~L. {Crawford} \& E.~R. {Craine},
  362, \dodoi{10.1117/12.176725}

\bibitem[{{Wittenmyer} {et~al.}(2009){Wittenmyer}, {Endl}, {Cochran},
  {Levison}, \& {Henry}}]{wittenmyer:2009}
{Wittenmyer}, R.~A., {Endl}, M., {Cochran}, W.~D., {Levison}, H.~F., \&
  {Henry}, G.~W. 2009, \apjs, 182, 97, \dodoi{10.1088/0067-0049/182/1/97}

\bibitem[{{Wittrock} {et~al.}(2016){Wittrock}, {Kane}, {Horch}, {Hirsch},
  {Howell}, {Ciardi}, {Everett}, \& {Teske}}]{wittrock:2016}
{Wittrock}, J.~M., {Kane}, S.~R., {Horch}, E.~P., {et~al.} 2016, \aj, 152, 149,
  \dodoi{10.3847/0004-6256/152/5/149}

\bibitem[{{Wolszczan} \& {Frail}(1992)}]{wolszczan:1992}
{Wolszczan}, A., \& {Frail}, D.~A. 1992, \nat, 355, 145,
  \dodoi{10.1038/355145a0}

\bibitem[{{Wright} \& {Eastman}(2014)}]{wright:2014}
{Wright}, J.~T., \& {Eastman}, J.~D. 2014, \pasp, 126, 838,
  \dodoi{10.1086/678541}

\bibitem[{{Zucker}(2003)}]{zucker:2003}
{Zucker}, S. 2003, \mnras, 342, 1291, \dodoi{10.1046/j.1365-8711.2003.06633.x}

\end{thebibliography}

\end{document}